\documentclass[aip,cha,preprint]{revtex4-1}
\usepackage{graphicx}
\usepackage{amsmath}
\usepackage{amssymb}
\usepackage{psfrag}
\usepackage{natbib}

\usepackage{color}

\newcommand{\pd}[2]{\displaystyle\frac{\partial #1}{\partial #2}}

\def \W {\Omega}

\def \vp {\varphi}

\def \ii {\text{i}}

\newcommand{\be}{\begin{equation}}
\newcommand{\ee}{\end{equation}}

\newcommand{\bea}{\begin{eqnarray}}
\newcommand{\eea}{\end{eqnarray}}

\begin{document}
	\title{Star-type oscillatory networks with generic Kuramoto-type coupling: a model for
``Japanese drums synchrony''}
	
	\author{Vladimir Vlasov}
	\affiliation{Institute for Physics and Astronomy, University of Potsdam,
14476 Potsdam, Germany}
	\author{Arkady Pikovsky}
	\affiliation{Institute for Physics and Astronomy, University of Potsdam,
14476 Potsdam, Germany}
\affiliation{Department of Control Theory, Nizhni Novgorod State University,
  Gagarin Av. 23, 606950, Nizhni Novgorod, Russia}
	\author{Elbert E. N. Macau}
\affiliation{National Institute for Space Research - INPE, 12227-010 
Sao Jose dos Campos, SP, Brazil}

	\date{\today}
	
	\begin{abstract}
	
	We analyze star-type networks of phase oscillators by virtue of two methods. For identical
	oscillators we adopt the Watanabe-Strogatz approach, that gives full analytical description of states, 
	rotating with constant frequency. For nonidentical oscillators, such states can be obtained
	by virtue of the self-consistent approach in a parametric form. In this case stability analysis cannot
	be performed, however with the help of direct numerical simulations we show which solutions are stable 
	and which not. We consider this system as a model for a drum orchestra, where we assume that
	the drummers follow the signal of the leader without listening to each other and the coupling parameters are
	determined by a geometrical organization of the orchestra.

	\end{abstract}

\maketitle

\begin{quotation}
	In the various studies of synchronization of globally coupled ensembles of oscillators, usually, it is assumed 
	that global coupling is produced by mean fields that act
	directly on each oscillator. However, in many natural systems (for example arrays of Josephson junctions),
	the global field is produced by a dynamical unit. In this paper we consider the case when this unit is a
	limit cycle oscillator, that can be described by a phase equation, similar to that for the other oscillators in the 
	ensemble. Such configuration of oscillators is often called star-type network. 
	In the first part of the paper we study the case of identical oscillators by virtue of the 
	Watanabe-Strogatz approach that gives full analytical analysis of steady solutions.
	In the second part 
	we examine inhomogeneous case, when the parameters of the coupling between each oscillator and the
	central element are different. Such network can be considered as a model for Japanese drums orchestra, 
	consisting of a battery of drummers and a leader, where the drummers only follow the signal from the leader.
	With the help of the self-consistent approach we find solutions when the global field rotates uniformly.
	These solutions are obtained semi-analytically in a parametric form. 
	In this work we analyze the case
	when the distribution of the parameters is determined by the geometric organization of the oscillators.
	However, this approach can be applied for an arbitrary distribution of the parameters.
	
\end{quotation}

%--------------------------------------------------------------------
\section{Introduction}

Studies of synchronization in populations of coupled oscillators attract high
interest. First, there are  
many experimental realizations of this effect, ranging from physical systems
(lasers, Josephson junctions, 
spin-torque oscillators) to biology (fireflies, genetically manipulated
circuits) and social 
activity of humans and animals (hand clapping, pedestrian footwalk on a bridge,
egg-laying in bird 
colonies), see reviews~\cite{Acebron-etal-05,Pikovsky-Rosenblum-15} for these
and other examples. 
Second, from the theoretical viewpoint, synchronization represents an example
of a nonequilibrium phase transition, and the challenging task is to describe it
as complete as 
possible in terms of suitable order parameters (global variables). In the
simplest setup of weakly 
coupled oscillators interacting via mean fields, the 
Kuramoto model of globally coupled phase
oscillators~\cite{Kuramoto-75,Kuramoto-84} 
and its generalizations are 
widely used. Here three main approaches have been developed. The original theory
by Kuramoto
is based on solving the self-consistent equations for the mean fields. In a
particular case of 
sine-coupled identical phase oscillators, the Watanabe-Strogatz 
theory~\cite{Watanabe-Strogatz-93,Watanabe-Strogatz-94} 
allows one to derive a closed set 
of equations for the order parameters. For non-identical sine-coupled
oscillators,  an important class 
of dynamical equations for the order parameters is obtained via the
Ott-Antonsen 
ansatz~\cite{Ott-Antonsen-08}.

Typically, coupling in the ensemble is considered as a force directly produced
by 
mean fields (either in a linear or nonlinear way). However, a more general setup
includes equations 
for global variables, driven by mean fields and acting on the oscillators. For
example, for Josephson 
junctions
and spin-torque oscillators, the coupling is due to a common 
load~\cite{Wiesenfeld-Colet-Strogatz-96,Pikovsky-13}, which may include
inertial 
elements (like capacitors and inductances) and therefore one gets an additional
system of 
equations (linear or nonlinear~\cite{Pikovsky-Rosenblum-09}, depending on the
properties of the load) for global variables and 
mean fields. A special situation appears when the mediator of the coupling is
not a linear damped 
oscillator (like an LCR load for Josephson junctions), but an active, limit cycle
oscillator. In this case the latter 
may be 
described by a phase equation similar to that describing one oscillator in the
population. Such an 
ensemble, where ``peripheral'' oscillators are coupled through one ``central''
oscillator, corresponds 
to a star-type network of
interactions~\cite{Kazanovich-Borisyuk-03,Burylko_etal-12,Kazanovich_etal-13}. 

One can make an analogy of such a network and a Japanese
drums orchestra.
 Such an orchestra consists of a battery of drummers and a leader, who sets the
rhythm of the music.
In our model the battery corresponds to an ensemble of oscillators coupled to a
leader, who also is an oscillator.
Although in practice each drummer represents a pulse oscillator (like a spiking
neuron), 
we adopt here the phase description like in the Kuramoto model. One should have in mind that
we do not aim in this paper to describe a real Japanese drum orchestra operation, but
rather use this the analogy to make the interpretation of the results demonstrative.

In this paper we apply the methods for description of global dynamics to such
star-type networks with
generic coupling (see \cite{Vlasov-Macau-Pikovsky-14} for generic algebraic
coupling), where interactions of oscillators with
the central element (i.e. their coupling constants and phase shifts) are in
general different, and are described by some joint distribution
function. 
We should stress that ``generic'' in this context means not including coupling
functions of arbitrary shape, but rather allowing for arbitrary distribution of parameters (amplitudes and phase shifts)
of the Kuramoto-type sin-coupling terms.
In the first part of the paper we present full analytical analysis of the
homogenous case (the intrinsic frequencies of all oscillators 
and the coupling constants are the same for all oscillators) in the framework of
the Watanabe-Strogatz (WS) approach (see~\cite{Vlasov-Zou-Pereira-15} for the
particular case of coupling coefficients). Due to 
the fact that the WS method leads to a
low-dimensional system of equations that describes dynamics of a homogenous
ensemble of any size, it is possible to 
analyze also stability of obtained solutions. We present solutions together with
their stability for different sets of 
parameters.
In the second part we perform the analysis of an inhomogeneous system with the
help of the self-consistent approach. After derivation of general equations
valid for an arbitrary distribution of coupling parameters, we consider a particular
example inspired by the analogy to the drum orchestra: we assume that the phase
shifts and coupling strengths follow from a geometric configuration of the ``battery''
and from the position of the ``conductor''.
In this case, the stability analysis could not be performed, but we
compare obtained self-consistent solutions with the results of direct numerical
simulations.

%-----------------------------------------------------
\section{The model}

	\begin{figure}[tbh]
		\includegraphics[width=0.5\columnwidth, clip]{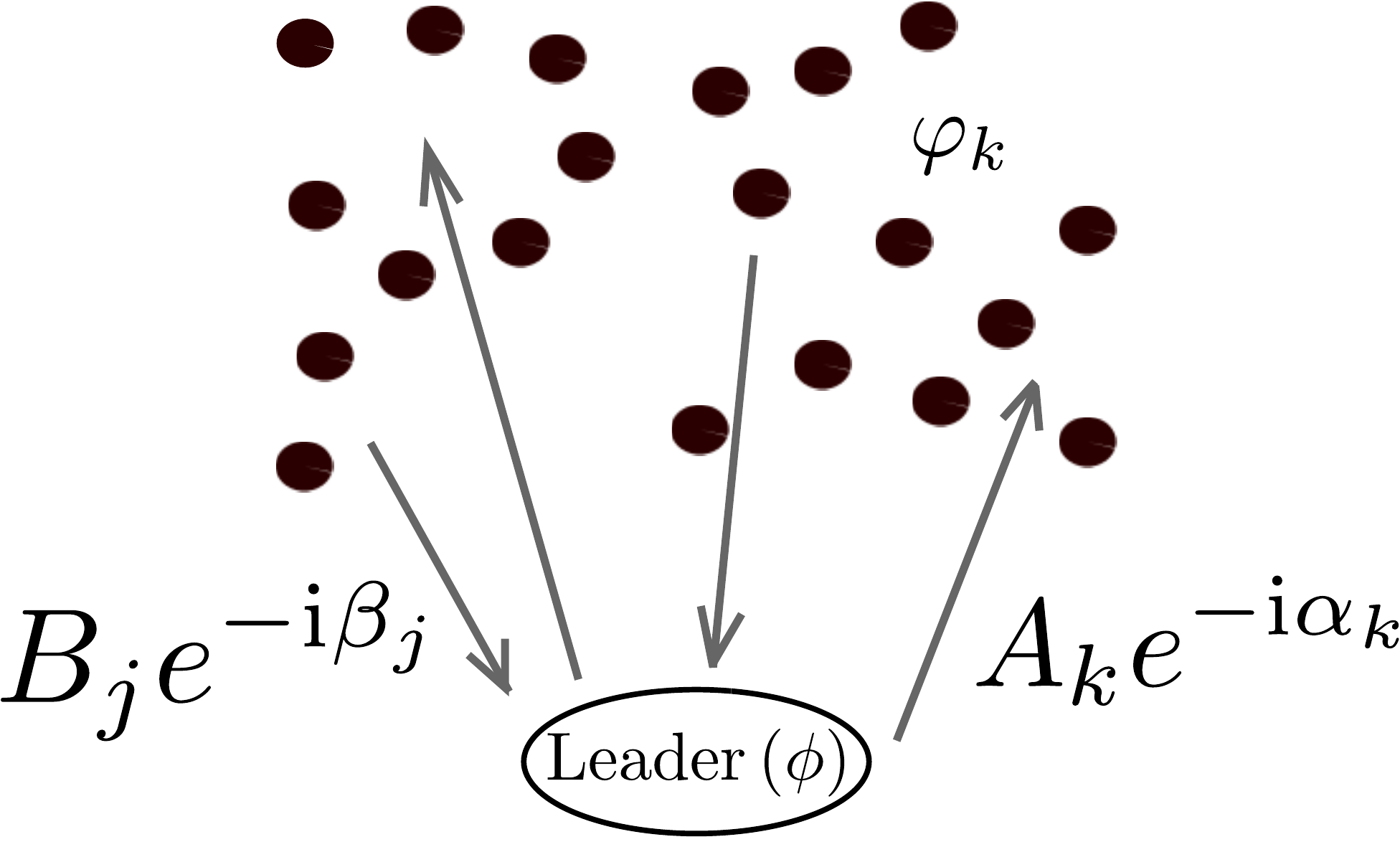}
		\caption{Configuration of the network, coupled through a
leader.}
		\label{fig.model-leader}
	\end{figure}

	We consider a system of phase oscillators with a leader-type coupling.
Such a network 
	structure~(Fig.~\ref{fig.model-leader}) is often called star network,
that is the simplest small-world network. In our setup, in 
	the most general case each phase oscillator $\vp_k$ has its own
frequency $\omega_k$ and is forced by the leader 
	oscillator (phase $\phi$) 
	with its own coupling strength $A_k$ and phase shift $\alpha_k$. At the
same time, the leader $\phi$ has its own 
	frequency $\omega_0$ and is forced by every other oscillator $\vp_j$
with coupling coefficient $B_j$ and 
	phase shift $\beta_j$. The dynamical equations thus read
	\be
		\label{gen.0}
		\begin{split}
			\dot{\vp_k}&=\omega_k+A_k\sin(\phi-\vp_k-\alpha_k),\quad
k=1\ldots N,\\
			\dot{\phi}&=\omega_0+{1\over N}\sum_{j=1}^N B_j
\sin(\vp_j-\beta_j-\phi).
		\end{split}
	\ee
	The system~(\ref{gen.0}) can be rewritten in terms of the mean field
	\be
		\label{gen.1}
		\begin{split}
			\dot{\vp_k}&=\omega_k+{\rm Im}(A_k e^{\ii
(\phi-\vp_k-\alpha_k)}),\\
			\dot{\phi}&=\omega_0+{\rm
Im}(\tilde{G}(t)e^{-\ii\phi}),\\
			\tilde{G}(t)&={1\over N}\sum_{j=1}^N
B_je^{\ii(\vp_j-\beta_j)}.
		\end{split}
	\ee
	It is convenient to perform a variable transformation to the phase
differences $\Delta\vp_k$ between the oscillators $\vp_k$ and the leader $\phi$,
taking also into account the phase shift $\alpha_k$:
	\be
		\label{var.1}
		\Delta\vp_k=\vp_k-\phi+\alpha_k.
	\ee
	Then, the equations for $\Delta\vp_k$ and $\phi$ are
	\be
		\label{gen.2}
		\begin{split}
			\frac{d\Delta\vp_k}{dt}&=-\frac{d\phi}{dt}+\omega_k+{\rm
Im}(A_ke^{-\ii \Delta\vp_k}),\\
			\frac{d\phi}{dt}&=\omega_0+{\rm Im}(G(t)),\\
			G(t)&={1\over N}\sum_{j=1}^N
B_je^{\ii(\Delta\vp_j-\alpha_j-\beta_j)}.
		\end{split}
	\ee	
	The expression for the leader dynamics can be directly inserted into the
equations for $\Delta\vp_k$, and thus we obtain the effective mean-field-coupled closed system
	\be
		\label{gen.3}
		\begin{split}
			\frac{d\Delta\vp_k}{dt}&=\omega_k-\omega_0-{\rm Im}(G(t))+{\rm
Im}(A_k e^{-\ii \Delta\vp_k}),\\
			G(t)&={1\over N}\sum_{j=1}^N
B_je^{\ii(\Delta\vp_j-\alpha_j-\beta_j)}.
		\end{split}
	\ee
	The system~(\ref{gen.3}) is equivalent to the phase
model~(\ref{gen.0}) for the leader-type (star) networks.
	This model in the form~(\ref{gen.3}) is similar to the models
of   Josephson junction arrays with the star-like topology, which have been considered previously
in~\cite{Tsang-Mirollo-Strogatz-Wiesenfeld-91,
Golomb-Hansel-Shraiman-Sompolinsky-92,Swift-Strogatz-Wiesenfeld-92}.  Analytical 
analysis similar to that described below has been performed for Josephson junction arrays
in~\cite{Vlasov-Pikovsky-13} .
In~\cite{Kazanovich-Borisyuk-03} the model~(\ref{gen.0}) for the systems with
center element (leader coupling) has been considered in the case of identical coupling
parameters and the distribution of natural frequencies of the leaf oscillators.
	
	In this work we are going to study the cases of identical and
nonidentical oscillators separately. Below we present the analytical analysis
for these two cases together with numerical simulations for a particular example
of nonidentical oscillators. In case of identical oscillators, we apply the
Watanabe-Strogatz approach. For the analysis of the case of nonidentical units
the self-consistent approach is used.

%--------------------------------------
\section{Identical oscillators}

	If all the parameters $\omega_k$, $A_k$, $B_j$, $\beta_j$ and $\alpha_j$
are identical, then the form of the phase equation (which is a particular case of Eq.~(\ref{gen.3}))
	\be
		\label{gen.3a}
		\begin{split}
			\frac{d\Delta\vp_k}{dt}&=\omega-\omega_0-{\rm
Im}\left(G(t)\right)+{\rm Im}(A e^{-\ii \Delta\vp_k}),\\
			G(t)&=Be^{-\ii(\alpha+\beta)}{1\over N}\sum_{j=1}^N
e^{\ii\Delta\vp_j}.
		\end{split}
	\ee
	allows us to use the WS
ansatz~\cite{Watanabe-Strogatz-93,Watanabe-Strogatz-94}, which is applicable to
any system of identical phase equations of the general form
	\be
		\label{WS.gen}
		\dot{\varphi_k}=f(t)+{\rm Im}\left(F(t)e^{-\ii\varphi_k}\right),
	\ee
	with arbitrary real  function $f(t)$ and complex function $F(t)$. It consists of the idea
that after WS variable transformation~(\ref{WS.1}), the dynamics of the
ensemble~(\ref{WS.gen}) is characterized by one global complex variable $z=z(t)$
and one real global variable $\Psi=\Psi(t)$, and $N$ constants of motion
$\psi_k$ (of which only $N-3$ are independent). We use the formulation of the WS
theory presented in~\cite{Pikovsky-Rosenblum-11}. WS transformation~(\ref{WS.1})
is 
	essentially the M\"obius
transformation~\cite{Marvel-Mirollo-Strogatz-09} in the form
	\be
		\label{WS.1}	
e^{\ii\varphi_k}=\frac{z+e^{\ii(\psi_k+\Psi)}}{1+z^*e^{\ii(\psi_k+\Psi)}},
	\ee
	with additional constraints $\sum_i e^{\ii\psi_i}=\sum_i\cos 2\psi_i=0$.
Then the global variables' dynamics is determined by
	\be
		\label{WS.glob.1}
		\begin{split}
			\dot{z}&=\ii f(t) z+ {F(t)\over 2}-{F^*(t)\over 2}z^2, \\
			\dot{\Psi}&=f(t) +{\rm Im}(z^* F(t)).
		\end{split}
	\ee
	
	Comparing the system~(\ref{gen.3a}) with~(\ref{WS.gen})  we see that in
our case $f(t)=\omega-\omega_0-{\rm Im}\left(G(t)\right)$ and $F(t)=A$. The next
step is to express the function $G(t)$, that is essentially the order parameter
multiplied by a complex number, in the new global variables. In general, such an
expression is rather complex (see~\cite{Pikovsky-Rosenblum-11} for details), but
in the thermodynamic limit $N\to\infty$ and for a uniform distribution of constants of
motion $\psi$ (the index has been dropped because constants now have a
continuous distribution) the order parameter is equal to $z$ {(see
Appendix~\ref{App:id.WS.z})}.
The constants $\psi$, as well as the WS variables $z(0)$ and $\Psi(0)$,
are determined by initial conditions
for the original phases $\Delta\varphi(0)$. In fact, any distribution of the constants 
is possible, and for each such distribution the dynamics will be different. 
However, as has been argued in Ref.~\cite{Pikovsky-Rosenblum-09}, 
in presence of small perturbations, initially non-uniform constants tend toward
the uniform distribution, which also the one appearing in the 
Ott-Antonsen ansatz~\cite{Ott-Antonsen-08}. Due to these
special relevance of the uniform distribution of the constants $\psi$,
we consider only this case below.

In this case, it follows  from (\ref{WS.glob.1}) that $\Psi$ does not enter the
equation for $z$, so we obtain a closed equation for $z$ that describes
system~(\ref{gen.3a}):
	\be
		\label{id.WS.1}
		\dot{z}=\ii\left(\Delta\omega-B\,{\rm
Im}(ze^{-\ii\delta})\right)z-A\frac{z^2-1}{2},
	\ee
	where $\Delta\omega=\omega-\omega_0$ and $\delta=\alpha+\beta$.

	For a further analysis, it is appropriate to represent the complex variable
$z=\rho e^{\ii\Delta\Phi}$ in polar form. Thus
	\be
		\label{id.WS.3}
		\begin{split}
			\frac{d\rho}{dt}&=A\frac{1-\rho^2}{2}\cos\Delta\Phi, \\		
\frac{d\Delta\Phi}{dt}&=\Delta\omega+B(\sin\delta)\rho\cos\Delta\Phi-\frac{
A+(A+2B\cos\delta)\rho^2}{2\rho}\sin\Delta\Phi.
		\end{split}
	\ee
	Note that Eqs.~(\ref{id.WS.3}) are invariant under the following
transformation of variables and parameters: $\Delta\Phi\to-\Delta\Phi$,
$\Delta\omega\to-\Delta\omega$ and $\delta\to-\delta$.
	
\subsection{Steady states}
	
	We start the analysis of~(\ref{id.WS.3}) with finding its steady states.
From the first equation in~(\ref{id.WS.3}) it follows that there are two types
of steady states with $\dot\rho=0$: synchronous with $\rho=1$ and asynchronous
with $\cos\Delta\Phi=0$. The synchronous steady state gives	
	\be
		\label{id.WS.rho=1.b}
		\begin{split}
			\rho&=1, \\	
\frac{d\Delta\Phi}{dt}&=\Delta\omega-\sqrt{A^2+B^2+2AB\cos\delta\,}
\sin\left(\Delta\Phi+\arcsin\frac{A+B\cos\delta}{\sqrt{A^2+B^2+2AB\cos\delta\,}}-\frac{\pi}{2}\right).
		\end{split}
	\ee
	From~(\ref{id.WS.rho=1.b}) it follows that the steady solution
$\frac{d\Delta\Phi}{dt}=0$ exists only if
$|\Delta\omega|\leq\sqrt{A^2+B^2+2AB\cos\delta\,}$.

	By rescaling time, we can reduce the number of parameters.
Eq.~(\ref{id.WS.rho=1.b}) suggests that the mostly convenient rescaling is 
	\be
		\label{id.WS.repar.t}
		t'=t\sqrt{A^2+B^2+2AB\cos\delta\,}.
	\ee
	This rescaling is quite general except for two special cases when
$\cos\delta=-1$ and $B=A$ (see Appendix~\ref{App:Special.case}) or $A=B=0$, the
latter case is just one of uniformly rotating uncoupled phase oscillators that
does not present any interest.
	So in the new parametrization Eqs.~(\ref{id.WS.rho=1.b}) have the form
	\be
		\label{id.WS.rho=1.c}
		\begin{split}
			\rho&=1, \\
			\frac{d\Delta\Phi}{dt}&=\Delta
x-\sin\left(\Delta\Phi+\xi-\frac{\pi}{2}\right),
		\end{split}
	\ee
	where
	\be
		\label{id.WS.repar.g.xi}
		\Delta x=\frac{\Delta\omega}{\sqrt{A^2+B^2+2AB\cos\delta\,}\,} \
\ \ \text{and} \ \ \
\sin\xi=\frac{A+B\cos\delta}{\sqrt{A^2+B^2+2AB\cos\delta\,}\,}.
	\ee 
	Thus the steady solutions of~Eq.~(\ref{id.WS.rho=1.c}) have the
following phases
	\be
		\label{id.WS.rho=1.fp}
		{\Delta\Phi_s}_1=\frac{\pi}{2}+\arcsin\Delta x -\xi, \ \ \
{\Delta\Phi_s}_2=-\frac{\pi}{2}-\arcsin\Delta x-\xi.
	\ee
	In the new parametrization Eqs.~(\ref{id.WS.3}) have the form
	\be
		\label{id.WS.4s}
		\begin{split}
			\frac{d\rho}{dt}&=g\frac{1-\rho^2}{2}\cos\Delta\Phi, \\
			\frac{d\Delta\Phi}{dt}&=\Delta
x+(\cos\xi)\rho\cos\Delta\Phi-\frac{g+(2\sin\xi-g)\rho^2}{2\rho}\sin\Delta\Phi,
		\end{split}
	\ee
	where $g=\frac{A}{\sqrt{A^2+B^2+2AB\cos\delta\,}\,}\geq0$.
	Note that similar to Eqs.~(\ref{id.WS.3}), Eqs.~(\ref{id.WS.4s}) are
invariant to the following transformation of variables and parameters
$\Delta\Phi\to-\Delta\Phi$, $\Delta x\to-\Delta x$ and $\cos\xi\to-\cos\xi$. Due
to this symmetry we can consider only the case when $\cos\xi\ge 0$.

	The asynchronous steady states can be found from
	\be
		\label{l.id.WS.DeltaPhi=pi/2.1}
		\begin{split}
			\Delta\Phi&=\pm\,\pi/2, \\
			0&=\Delta x\mp\frac{g+(2\sin\xi-g)\rho^2}{2\rho}.
		\end{split}
	\ee
	Eq~(\ref{l.id.WS.DeltaPhi=pi/2.1}) gives two asynchronous steady
solutions:
	\be
		\label{l.id.WS.DeltaPhi=pi/2.3.sol}
		{z_a}_{1,2}=\ii\frac{\Delta x\pm\sqrt{\Delta
x^2-g(2\sin\xi-g)}}{2\sin\xi-g}.
	\ee
	It is convenient to rewrite Eq.~(\ref{l.id.WS.DeltaPhi=pi/2.3.sol}) as
	\be
		\label{l.id.WS.DeltaPhi=pi/2.3.sol.1}
		{z_a}_{1,2}=\text{sign}(\Delta x)\,\ii\frac{|\Delta
x|\pm\sqrt{\Delta x^2-g(2\sin\xi-g)}}{2\sin\xi-g}.
	\ee
	Note, that here the cases split, depending on the value of
$(2\sin\xi-g)$. If $|2\sin\xi-g|>g$, then, because $\rho=|z|\le1$, 
the solution ${z_a}_{1}$ exists only if $|\Delta x|\leq|\sin\xi|$. If $|2\sin\xi-g|\le g$,
then, also because $\rho=|z|\le1$,  the  solution ${z_a}_{2}$ exists only if $|\Delta
x|\geq\sin\xi$. If $2\sin\xi-g=0$, then the asynchronous steady solutions are
	\be
		\label{l.id.WS.DeltaPhi=pi/2.3.sol.2}
		{z_a}_{1,2}=\pm\,\text{sign}(\Delta x)\,\ii\frac{g}{2|\Delta
x|},
	\ee
	but the condition on $|\Delta x|\ge\sin\xi=g/2$ is still the same.
	
	Note that the expression $2\sin\xi-g$ is equal to
$\frac{A+2B\cos\delta}{\sqrt{A^2+B^2+2AB\cos\delta\,}\,}$;  if $\cos\delta\geq0$
this expression is always positive, so that $g<1$ and $\sin\xi>g$. If $\cos\delta<0$, then
the sign of this expression depends on the sign of $A+2B\cos\delta$, but
$\sin\xi<g$.

%-----------------------------------------------------------------

\subsection{Stability analysis}
	
	In order to analyze stability of the asynchronous steady
solutions~(\ref{l.id.WS.DeltaPhi=pi/2.3.sol.1}) we linearize the system around
corresponding fixed point. The linearized system reads
	\be
		\label{l.id.WS.stab.asyn}
		\begin{split}
			\dot{a}_{1,2}&=\text{sign}(\Delta
x)\left(-(\cos\xi)\,\frac{|\Delta x|\pm\sqrt{\Delta
x^2-g(2\sin\xi-g)}}{2\sin\xi-g}\,a_{1,2}\pm\,\sqrt{\Delta
x^2-g(2\sin\xi-g)}\,b_{1,2}\right),\\
			\dot{b}_{1,2}&=\text{sign}(\Delta x)\left(|\Delta
x|-(\sin\xi)\,\frac{|\Delta x|\pm\sqrt{\Delta
x^2-g(2\sin\xi-g)}\,}{2\sin\xi-g}\right)a_{1,2},
		\end{split}
	\ee
	where $a_{1,2}$ and $b_{1,2}$ are real and imaginary parts of the perturbation, respectively.
	Although it is difficult to find explicit expressions for
the eigenvalues of the linear 
	system~(\ref{l.id.WS.stab.asyn}),
	{compared to a straightforward calculation of eigenvalues for the synchronous fixed points below,}  it is possible to find regions of
parameters where they are positive or negative, what is sufficient for determining stability of
asynchronous solutions {(see Appendix~\ref{App:id.sol} for detailed description of stability properties of the asynchronous states)}.
	
	There is one truly remarkable case when $\cos\xi=0$ or $\sin\delta=0$.
In this case, eigenvalues of the linear system~(\ref{l.id.WS.stab.asyn}) for
${z_a}_{2}$ are purely imaginary, what opens a possibility for the fixed point
${z_a}_{2}$ to be neutrally stable (see subsection~\ref{sec:rev} for details).

	For two synchronous fixed points~(\ref{id.WS.rho=1.fp}):
${z_s}_{1}=e^{\ii{\Delta\Phi_s}_1}$ and ${z_s}_{2}=e^{\ii{\Delta\Phi_s}_2}$, the
corresponding linearized system reads
	\be
		\label{l.id.WS.stab.syn}
		\begin{split}
			\dot{a}_{1,2}=&\Biggl[\mp\sqrt{1-\Delta
x^2}+(\sin\xi-g)(-\Delta x\cos\xi\pm\sqrt{1-\Delta x^2}\sin\xi)\Biggr]a_{1,2}+\\
			&+\Biggl[(\sin\xi-g)(\pm\sqrt{1-\Delta
x^2}\cos\xi+\Delta x\sin\xi)\Biggr]b_{1,2},\\
			\dot{b}_{1,2}=&\Biggl[\cos\xi(-\Delta
x\cos\xi\pm\sqrt{1-\Delta x^2}\sin\xi)\Biggr]a_{1,2}+\\
			&+\Biggl[-\sin\xi(-\Delta x\cos\xi\pm\sqrt{1-\Delta
x^2}\sin\xi)\Biggr]b_{1,2},
		\end{split}
	\ee
	with the same meaning of linear perturbations $a_{1,2}$, $b_{1,2}$.
	Linear system~(\ref{l.id.WS.stab.syn}) has two eigenvalues:
	\be
		\label{l.id.WS.stab.syn.eig}
		\begin{split}
			{\lambda_s}_{1,2}^{1}&=g(\Delta
x\cos\xi\mp\sqrt{1-\Delta x^2}\sin\xi),\\
			{\lambda_s}_{1,2}^{2}&=\mp\sqrt{1-\Delta x^2}.
		\end{split}
	\ee
	Their signs depend on the values of the parameters. We will outline all
possible steady solutions together with their stability below {(see Appendix~\ref{App:id.sol} for a full description of stability properties of the steady solutions)}.
%--------------------------------------------------------------------
\subsection{Reversible case when condition $\sin\delta=0$ holds}
\label{sec:rev}

\begin{figure}[tbh]
		\includegraphics[width=0.6\columnwidth, clip]{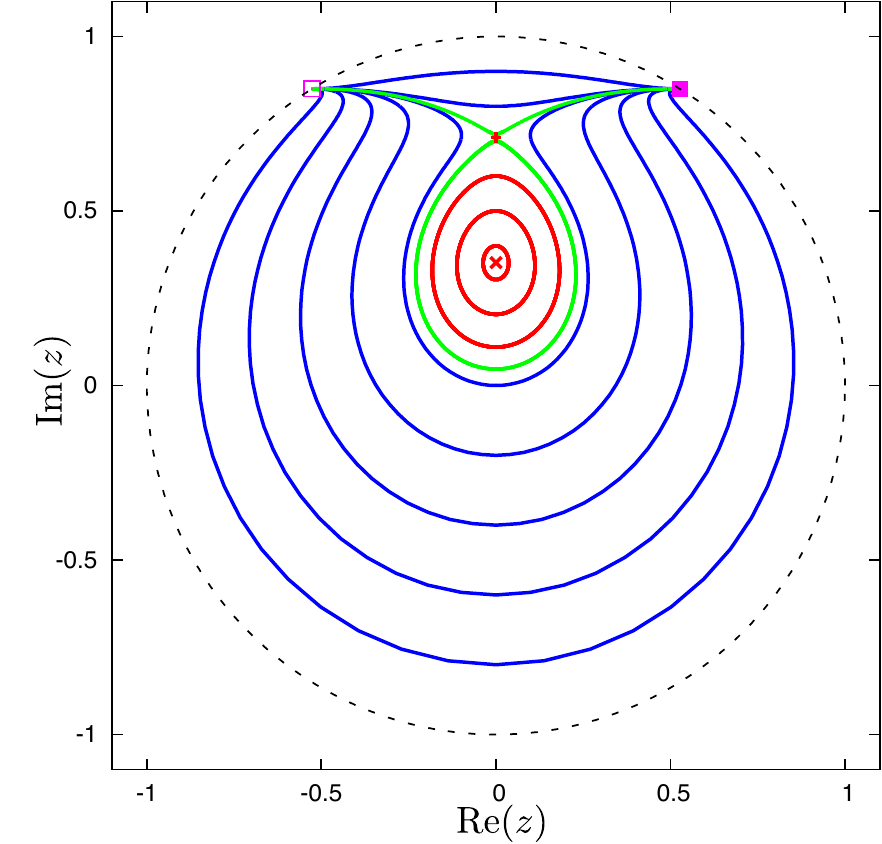}
		\caption{The phase portrait for the reversible order parameter
dynamics (case $\sin\xi=1$, $\Delta x=0.85$, and $g=0.4$).  
 Trajectories connecting the unstable synchronous fixed point (denoted as the
empty square) and the stable one (filled square) are shown in blue. Trajectories
oscillating around neutrally stable fixed point (x-cross) are shown in red.
Stable and unstable manifolds of the saddle (denoted as the plus marker) are
shown in green.}
		\label{fig.phase-diag-0}
	\end{figure}

	As has been shown by the stability analysis above, when $\sin\delta=0$, the
steady state
	${z_a}_{2}$ is neutrally stable for any $\Delta x$. The neutral
stability can be proved by the fact that when $\sin\delta=0$ (and, as a consequence, $\cos\xi=0$),
Eq.~(\ref{id.WS.1}) is invariant to the  transformation ${\rm Im}(z)\to{\rm Im}(z)$, ${\rm Re}(z)\to
-{\rm Re}(z)$, when the time direction is also changed: $t\to -t$.
This transformation, which leaves the imaginary axis invariant, is an involution. 
	Thus,  the dynamics is reversible~\cite{Roberts-Quispel-92}: any trajectory
that crosses ${\rm Im}(z)$ axes twice is a neutrally stable closed curve
(Fig.~\ref{fig.phase-diag-0}). This set of quasi-Hamiltonian trajectories
(surrounded by a homoclinic one) coexists with an attractor and a repeller,
which are symmetric to each other (for other examples of coexistence
of conservative and dissipative dynamics in reversible systems see, e.g., \cite{Politi-Oppo-Badii-86}).  
In direct numerical simulations, one
observes, depending on initial conditions, either a synchronous state, or
oscillations of the order parameter $z$. 
	
%-----------------------------------------------------------------
\subsection{Synchronization scenarios}	
	
There are three main regions of parameters with three different transitions from
asynchronous steady solution to 
synchronous one.

\begin{figure}[tbh]
		\includegraphics[width=0.49\columnwidth, clip]{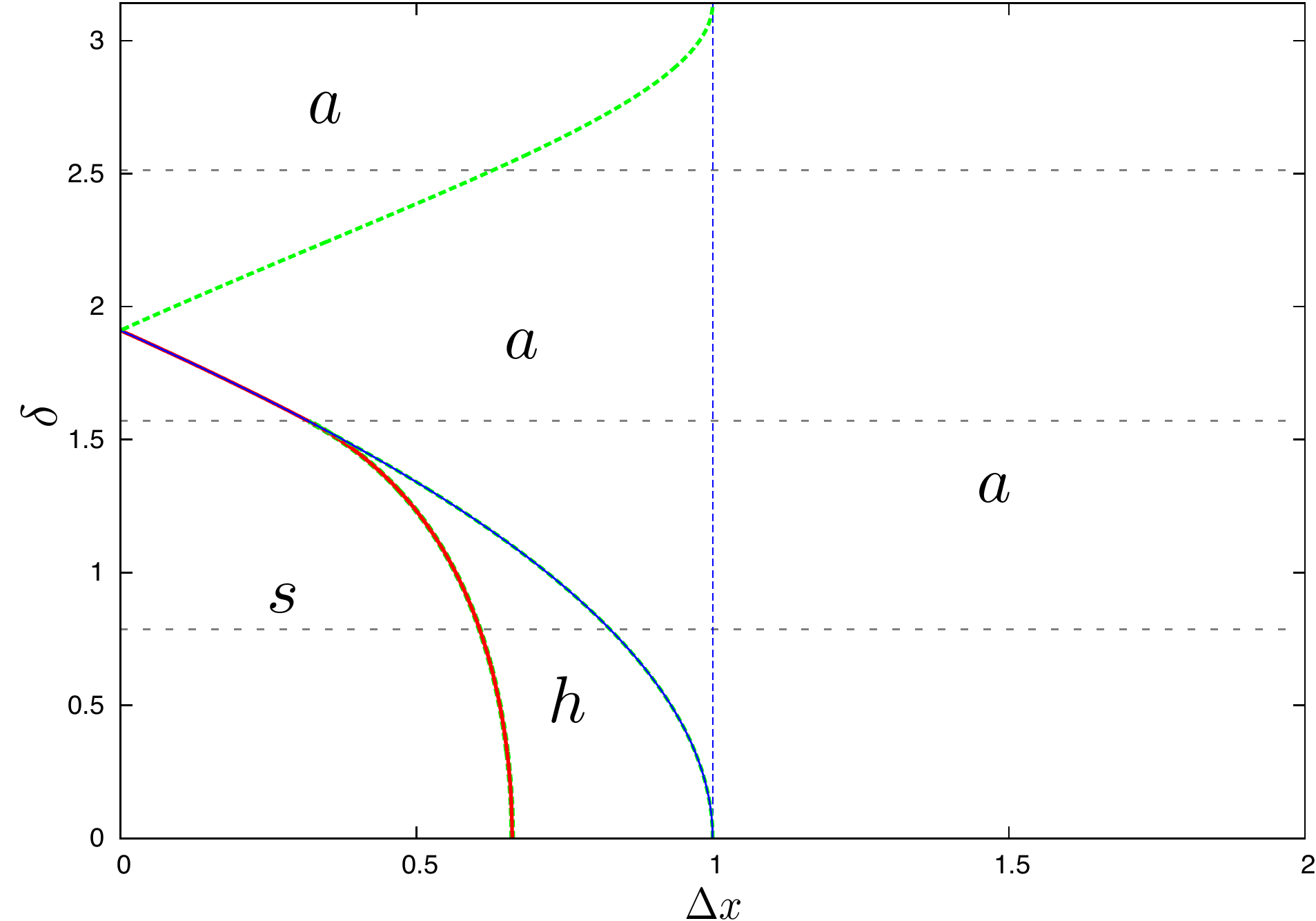}
		\caption{Different regimes in the parameter plane $(\delta,\Delta x)$ for $A=1$ and $B=3$.
		The area, where the synchronous steady solution is stable, is denoted with $s$,
		the areas, where asynchronous steady solution is stable, are denoted with $a$,
		the area of the hysteresis is denoted with $h$.
		Solid red line is stability line of the asynchronous state,
		solid blue line is stability line of the synchronous state.
		Dashed green line denotes the area of existence of unstable asynchronous solution and
		dashed blue line - the area of existence of unstable synchronous solution.
		Horizontal dashed gray lines are cuts of the diagram illustrated in Fig~\ref{fig.st-dr-123}
		((a) corresponds to the bottom line, (b) - to the middle line and (c) - to the top line).}
		\label{fig.st-dr-diag}
	\end{figure}

\begin{figure}[tbh]	
	\centering
	(a) \includegraphics[width=0.3\columnwidth, clip]{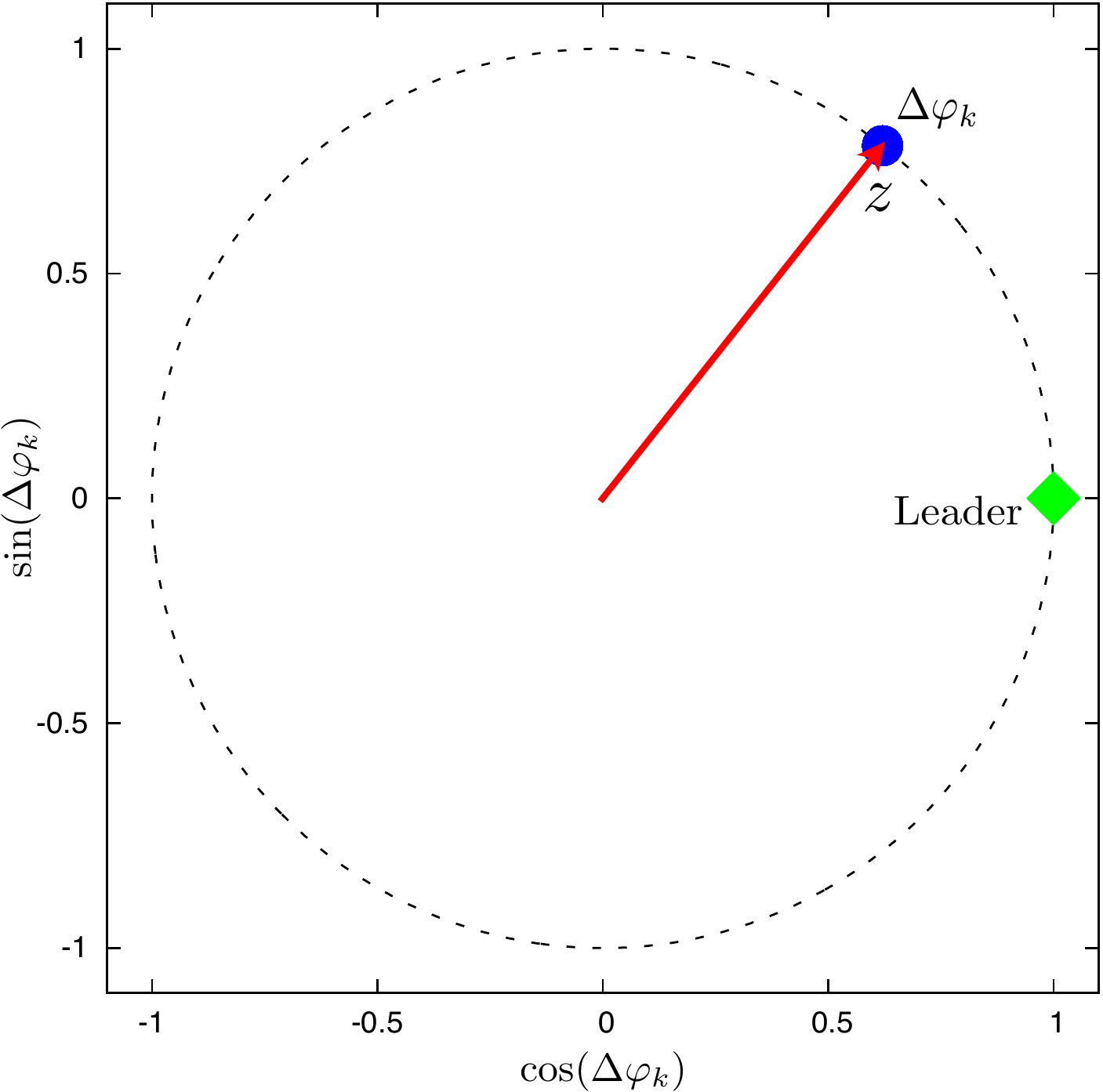}\hfill
		(b) \includegraphics[width=0.3\columnwidth, clip]{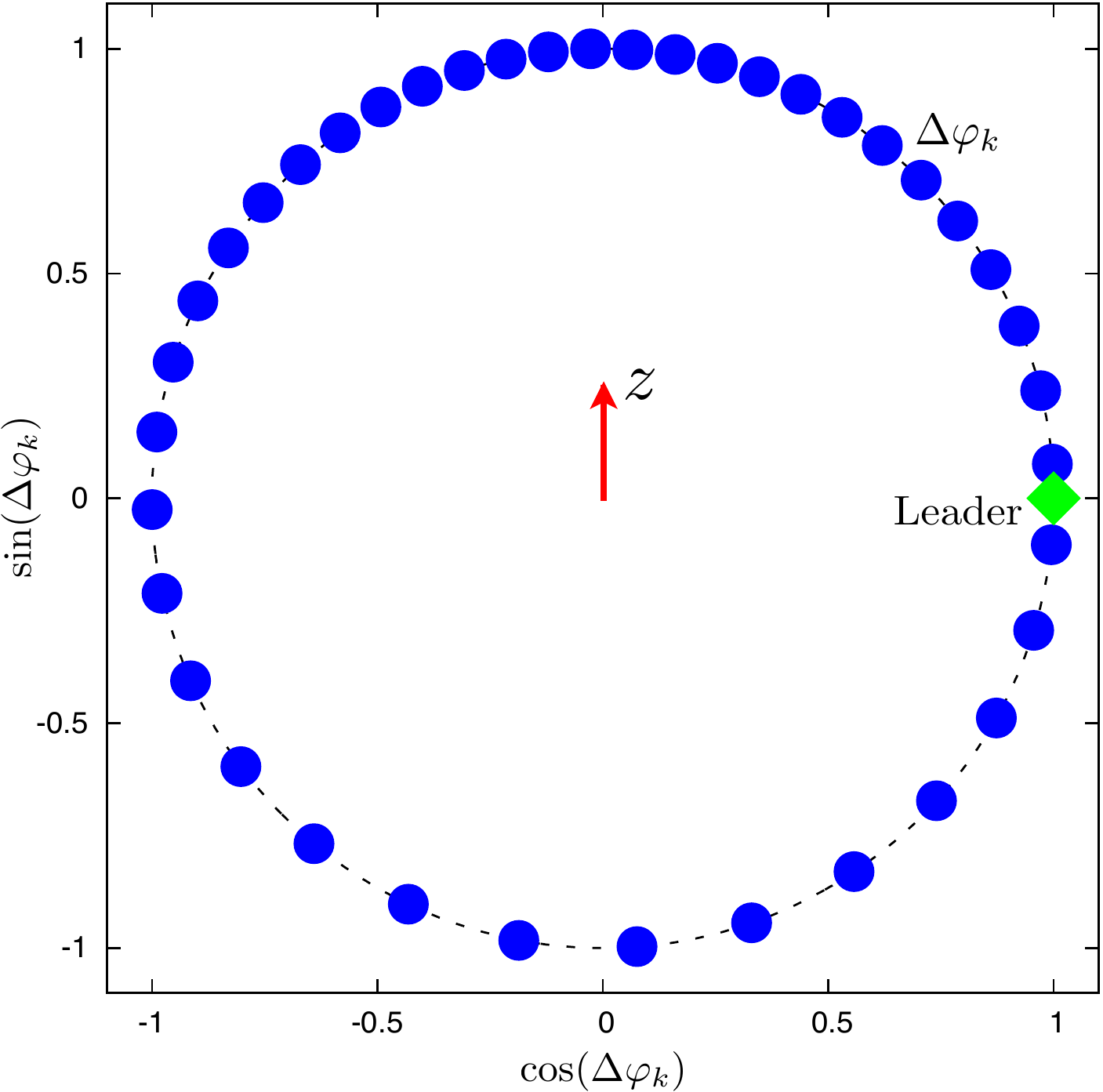}
		\caption{Snapshots of the oscillators for synchronous ((a) $\Delta x = 0.3$)
		and asynchronous ((b) $\Delta x = 0.7$) cases.
		The parameters are the same as for the Fig.~\ref{fig.st-dr-123}(a).}
		\label{fig.st-dr-phases}
	\end{figure}

We present the diagram of different states in the parameter plane $(\delta,\Delta x)$
in the Fig.~\ref{fig.st-dr-diag}. There are three main domains: the domain, where 
the synchronous solution (illustrated in  Fig.~\ref{fig.st-dr-phases}(a)) is stable;
the domain where the asynchronous solution 
(illustrated in Fig.~\ref{fig.st-dr-phases}(b)) is stable; and the domain of bistability. 
As parameters $\delta$ and $\Delta x$ describe two main properties of the star-type
coupling, namely the phase shift in the coupling and the frequency mismatch between the central and peripheral
elements, correspondingly, the interpretation of these domains is straightforward. Phase shifts close to zero, as well as small frequency mismatches facilitate synchrony; bistability is observed when the mismatch is relatively large while the phase shift $\delta$ is close to the optimal one for synchronization (zero).

We show the transitions in dependence  on the
relative frequency mismatch $\Delta x$ 
between the  oscillator's natural frequency and the frequency of the leader. 
Stability of the solutions depends on the sign of the
frequency mismatch $\Delta x$. A detailed description of the solutions we
present in Appendix~\ref{App:id.sol}.

	\begin{figure}[tbh]
		\centerline{(a)\includegraphics[width=0.4\columnwidth, clip]{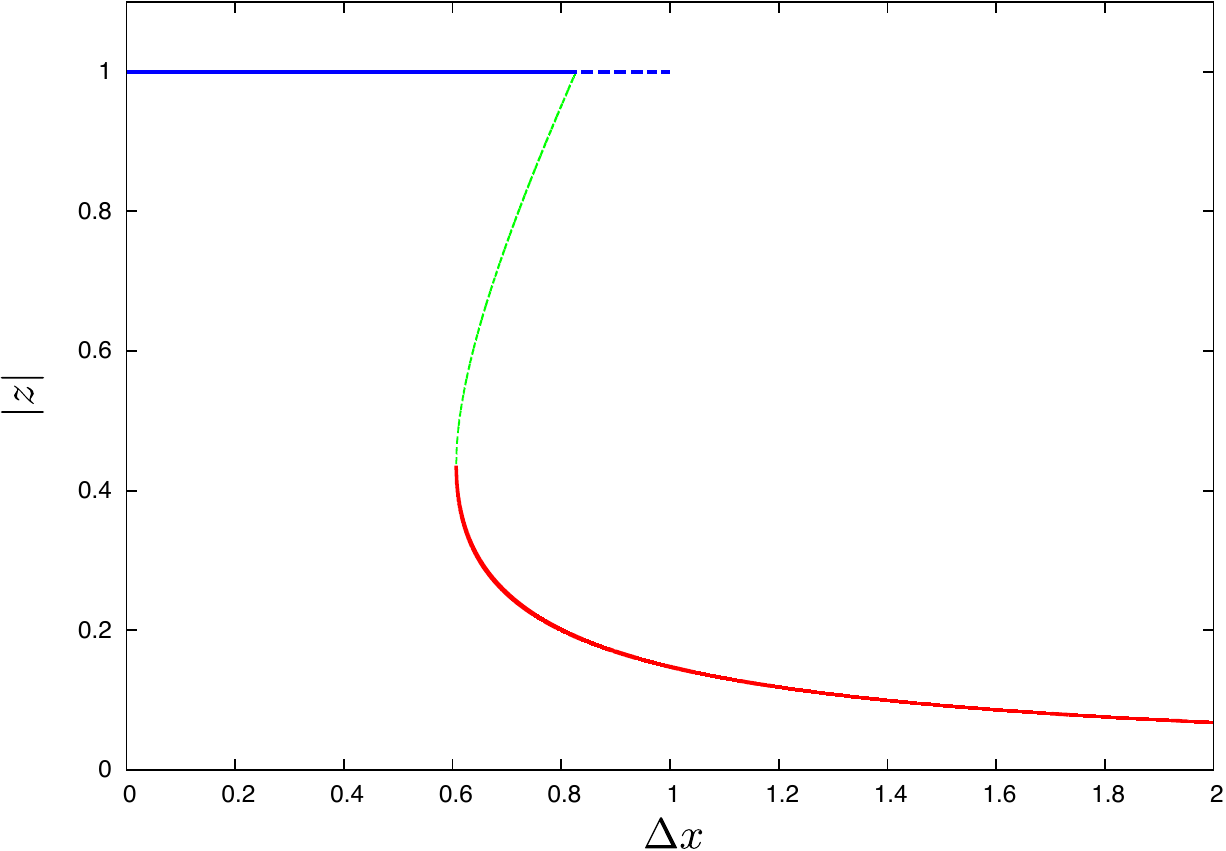}
		(b)\includegraphics[width=0.4\columnwidth, clip]{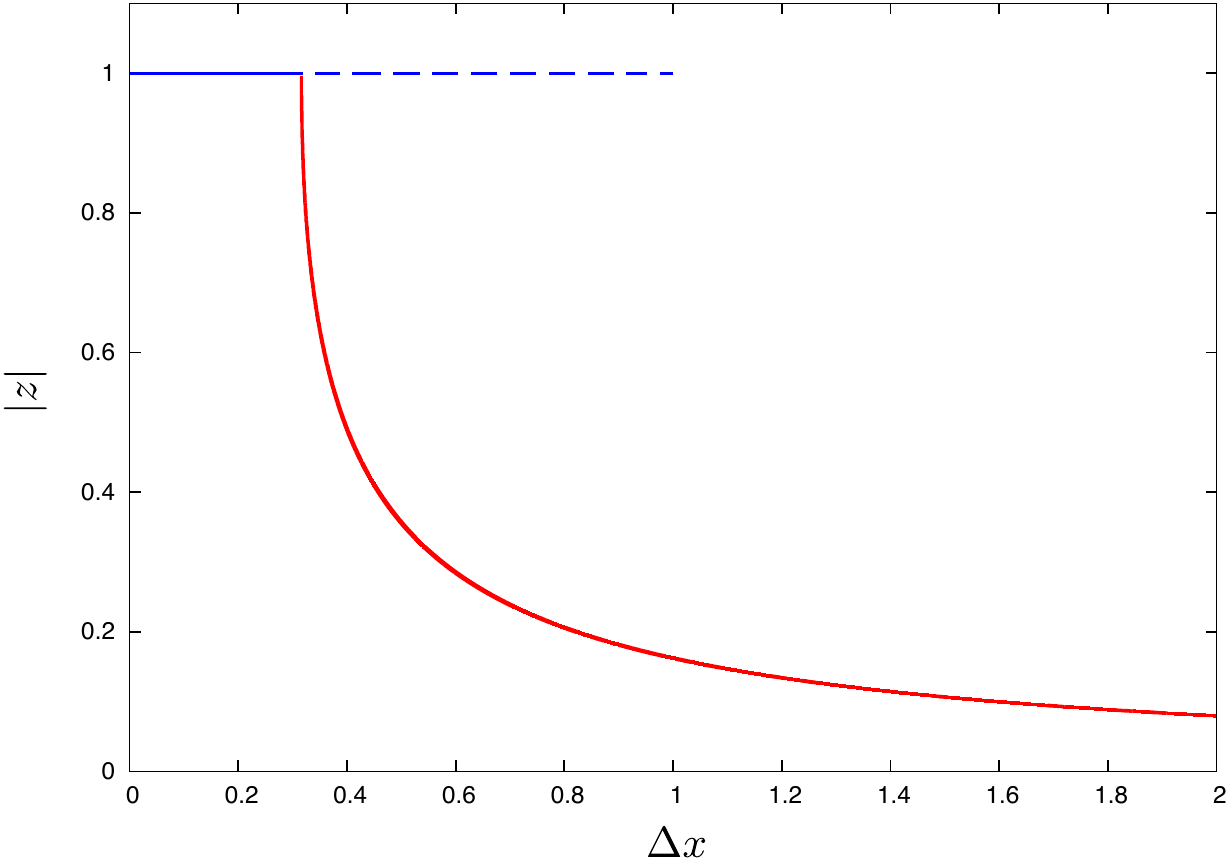}}
		(c)\includegraphics[width=0.4\columnwidth, clip]{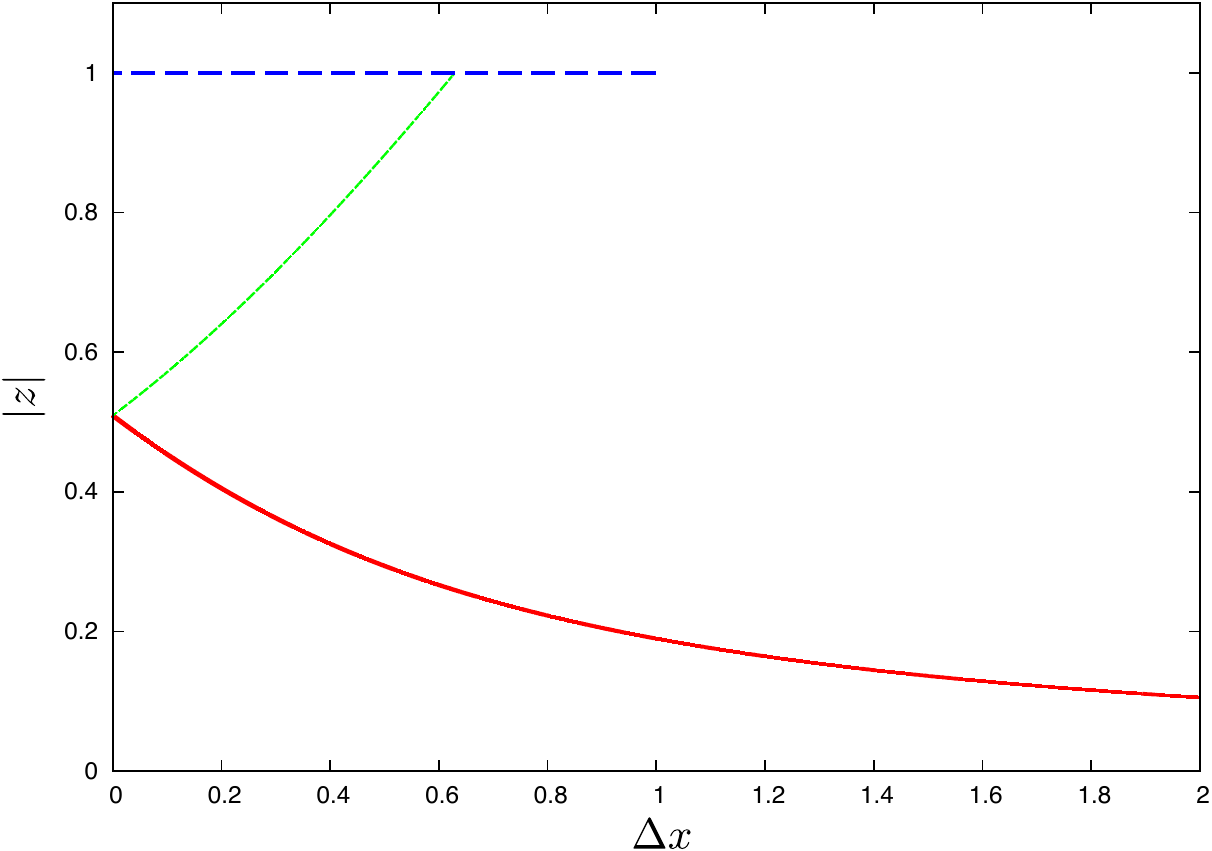}
		\caption{The dependence of the order parameter $|z|$ on the
relative frequency mismatch $\Delta x$, for the case $\sin\xi> g\geq 0$ (a),
for the case $g\geq\sin\xi\geq0$ (b) and for the case $\sin\xi<0$ (c). The red
curve denotes stable asynchronous steady solution, the green curve denotes
unstable asynchronous steady solution and solid blue line denotes synchronous
steady solution, while dashed blue line denotes unstable synchronous steady
solution.}
		\label{fig.st-dr-123}
	\end{figure}
	
\paragraph{}
In the first case~(Fig.~\ref{fig.st-dr-123}(a)) there is a hysteretic synchronization transition. 
If $\Delta x>0$, there is a stable asynchronous steady solution that exists for
large absolute values of $|\Delta x|$, and 
a stable synchronous steady solution that exists for small $|\Delta x|$. These
solutions coexist for a bounded region 
of  $|\Delta x|$, thus forming hysteresis. If the frequency mismatch is negative
$\Delta x<0$, the synchronous steady solution 
is still stable, and also the synchronous limit cycle solution 
becomes stable for large $|\Delta x|$, while the asynchronous steady solution
becomes unstable.

\paragraph{}
In the second case~(Fig.~\ref{fig.st-dr-123}(b)) there is no hysteresis. In this case
there is only one asynchronous steady solution 
existing for large values of $|\Delta x|$ that is stable if $\Delta x>0$ and
unstable if $\Delta x<0$. If $\Delta x>0$,
the synchronous solution is stable for small values of $|\Delta x|$, while if
$\Delta x<0$ both synchronous (steady and limit cycle) solutions are stable for
all $|\Delta x|$.
	
\paragraph{}
In the third case~(Fig.~\ref{fig.st-dr-123}(c)) the transition is not hysteretic. For
$\Delta x>0$ there is only one stable steady 
solution that is the asynchronous one. And for $\Delta x<0$, for small $|\Delta x|$
there is stable asynchronous steady solution 
that transforms to synchronous steady solution for larger $|\Delta x|$, which 
with further increase of $|\Delta x|$ becomes 
a stable synchronous limit cycle.

%-------------------------------------------------------------------
\subsection{Star-like and mean field coupling}

The analytical approach described above can be partially applied for the system
of identical oscillators coupled not only 
through interactions with a leader, but also via a Kuramoto-Sakaguchi mean
field.
Adding such a mean-field means that we add additional all-to-all coupling between the leaf oscillators, thus
the original system~(\ref{gen.0}) reads
	\be
		\label{m-f.gen.0}
		\begin{split}
			\dot{\vp_k}&=\omega+A\sin(\phi-\vp_k-\alpha)+{1\over N}\sum_{j=1}^NC\sin(\vp_j-\vp_k-\gamma),\quad
k=1\ldots N,\\
			\dot{\phi}&=\omega_0+{1\over N}\sum_{j=1}^N B
\sin(\vp_j-\beta-\phi).
		\end{split}
	\ee
Parameter $C$ here describes relative strength of the direct mean-field coupling compared to that via the leader.
In the context of drum orchestra, this corresponds to a situation, where the drummers follow not only the leader, but
react on the other drummers. In a more general context, Eqs.~\eqref{m-f.gen.0} describe coupling via two 
mean field
channels, one direct (Kuramoto-Sakaguchi term), and one mediated by the leader.

The system~(\ref{m-f.gen.0}) can be similarly rewritten to the form of the system~(\ref{gen.3a}) with additional mean field $H(t)$:

%
%The system~(\ref{gen.3a}) with additional mean field $H(t)$ reads
	\be
		\label{m-f.gen.3a}
		\begin{split}
			\frac{d\Delta\vp_k}{dt}&=\omega-\omega_0-{\rm
Im}\left(G(t)\right)+{\rm Im}(A e^{-\ii \Delta\vp_k})+{\rm Im} (H(t)
e^{-\ii\Delta\vp_k}),\\
			G(t)&=Be^{-\ii(\alpha+\beta)}{1\over N}\sum_{j=1}^N
e^{\ii\Delta\vp_j}, \\
			H(t)&=Ce^{-\ii\gamma}{1\over N}\sum_{j=1}^N
e^{\ii\Delta\vp_j}.
		\end{split}
	\ee
	
The WS approach can be also applied to the new system~(\ref{m-f.gen.3a}), so
that according to~(\ref{WS.glob.1})
we obtain the following equation for the order parameter
	 \be
		\label{m-f.id.WS.1}
		\dot{z}=\ii(\Delta\omega-B\,{\rm
Im}(ze^{-\ii\delta}))z-A\frac{z^2-1}{2}+\frac{C}{2}(e^{-\ii\gamma}-e^{\ii\gamma}
|z|^2)z.
	\ee
Then we perform a similar analysis, which involves the same rescaling of
time~(\ref{id.WS.repar.t}) and 
reparameterization~(\ref{id.WS.repar.g.xi}) as in the previous case. In the rescaled time 
Eq.~(\ref{m-f.id.WS.1}) for the magnitude and the argument of the order parameter reads
	\be
		\label{m-f.id.WS.4s}
		\begin{split}		
\frac{d\rho}{dt}&=\frac{1-\rho^2}{2}(g\cos\Delta\Phi+q\cos\gamma\,\rho), \\
			\frac{d\Delta\Phi}{dt}&=\Delta
x-q\frac{1+\rho^2}{2}\sin\gamma+\cos\xi
\rho\cos\Delta\Phi-\frac{g+(2\sin\xi-g)\rho^2}{2\rho}\sin\Delta\Phi,
		\end{split}
	\ee
	where $q=\frac{C}{\sqrt{B^2+A^2+2BA\cos\delta\,}\,}\geq0$. Note that
similar to Eqs.~(\ref{id.WS.3}), Eqs.~(\ref{m-f.id.WS.4s}) are invariant to
the following transformation of variables and parameters
$\Delta\Phi\to-\Delta\Phi$, $\Delta x\to-\Delta x$ and $\cos\xi\to-\cos\xi$,
$\gamma\to-\gamma$. Thus, as above, we will consider only the case when
$\cos\xi\ge 0$.
	
	The synchronous steady solutions with $\rho=|z|=1$
of~Eq.~(\ref{m-f.id.WS.4s}) are
	\be
		\label{m-f.id.WS.rho=1.fp}
		{\Delta\Phi_s}_1=\frac{\pi}{2}+\arcsin(\Delta x-q\sin\gamma)
-\xi, \ \ \ {\Delta\Phi_s}_2=-\frac{\pi}{2}-\arcsin(\Delta x-q\sin\gamma) -\xi.
	\ee
The incoherent steady solutions should be found from the following equations
	\be
		\label{m-f.id.WS.DeltaPhi=pi/2.1}
		\begin{split}
			\cos\Delta\Phi&=-\frac{q\cos\gamma\,\rho}{g}, \\
			0&=\Delta
x-q\frac{1+\rho^2}{2}\sin\gamma-(\cos\xi)\rho\,\frac{q\cos\gamma\,\rho}{g}
\mp\frac{g+(2\sin\xi-g)\rho^2}{2\rho}\,\sqrt{1-\left(\frac{q\cos\gamma\,\rho}{g}
\right)^2}.
		\end{split}
	\ee

The system of equations~(\ref{m-f.id.WS.DeltaPhi=pi/2.1}) for $\rho$ and
$\Delta\Phi$ is rather complex for an analytical 
analysis, but it is clear that there are two main limiting cases. The first is
the case with large $C$ (which corresponds to the parameter $q$), this 
means that the dynamics of the system is mostly influenced by the Kuramoto mean field.
This case qualitatively coincides with the well 
studied case when $B=A=0$ with two synchronous fixed points (one stable and one
unstable with $|z|=1$) and one 
asynchronous fixed point (stability of which depends on the coupling parameters
and frequency mismatch). The second 
case is when the influence of the mean field is relatively small, what happens if  the
coupling strength $C$ (or $q$) is small. The qualitative 
picture for this case coincides with the limit $C=q=0$ considered in the main
part of this section. The quantitative results can 
be obtained by solving system~(\ref{m-f.id.WS.DeltaPhi=pi/2.1}) numerically. Note,
that our approach is still useful here 
because the numerical analysis of the reduced 
system~(\ref{m-f.id.WS.4s}) is much simpler then the original one.

%----------------------------------------------------
\section{Nonidentical oscillators}

Let us return to the original formulation of the problem with a generic
distribution of the coupling constants and 
phase shifts~(\ref{gen.3})
	\be
		\label{dr.gen.3}
		\begin{split}
			\frac{d\Delta\vp_k}{dt}&=\omega_k-\omega_0-{\rm Im}(G(t))+{\rm
Im}(A_k e^{-\ii \Delta\vp_k}),\\
			G(t)&={1\over N}\sum_{j=1}^N
B_je^{\ii(\Delta\vp_j-\alpha_j-\beta_j)},
		\end{split}
	\ee
	where dynamics of the leader 
	\be
		\label{dr.leader}
		\dot{\phi}=\omega_0+{\rm Im}(G(t))
	\ee
	does not enter in the equations for the phase differences.

\subsection{Self-consistent approach}
	We analyze the solutions of~(\ref{dr.gen.3}) in the thermodynamic limit
$N\rightarrow \infty$, where in this case the parameters $\omega$, $A$, $B$,
$\alpha$ and $\beta$ have a joint distribution density $w(p)=w(\omega, A, B,
\alpha, \beta)$, where $p$ is a general vector of parameters. Introducing the
conditional probability density function for the distribution of the phases at a given set $p$:
$W(\Delta\vp ,t \,|\, p)$, we can
rewrite system~(\ref{dr.gen.3}) as
	\be
		\label{s-c.eq}
		\begin{split}	
\frac{d\Delta\vp}{dt}=&\,\omega-\omega_0-Q\sin\Delta\Theta-A\sin\Delta\vp,\\
			G(t)=Q e^{\ii \Delta\Theta}&=\int
w(p)Be^{-\ii(\alpha+\beta)}\int_0^{2\pi}W(\Delta\vp,t\,|\,p)\,e^{\ii\Delta\vp}
d\Delta\vp \,dp,
		\end{split}
	\ee
	where $W(\Delta\vp,t\,|\,p)$ should be calculated from the Liouville
equation
	\be
		\label{s-c.FP}	
\pd{W}{t}+\pd{}{\Delta\vp}\left(\left[\omega-\omega_0-Q\sin\Delta\Theta-
A\sin(\Delta\vp)\right]W \right)=0.
	\ee
	
	Then, we look for a stationary solution for the distribution of the phase difference
$\Delta\vp$
	\be
		\label{s-c.cond}
		\pd{}{t}{W}(\Delta\vp,t\,|\,p)=0.
	\ee
	Stationarity of  the distribution of $\Delta\vp$ means that we are looking for the solutions
of the original system~(\ref{gen.0}), for which phases $\varphi$ rotate with a constant
frequency $\Omega$, where $\Omega$ denotes the frequency of the leader
(resulting from~\eqref{dr.leader})
	\be
		\label{s-c.leader}
		\Omega=\dot{\phi}=\omega_0+Q\sin\Delta\Theta.
	\ee
Then it is convenient to treat the unknowns $Q$, $\Delta\Theta$ and the
parameter $\omega_0$ as functions of $\Omega$. 
		
The stationary solution of the stationary Liouville equation~(\ref{s-c.FP}), is
either a delta-function or a
continuous distribution:
	\be
		\label{s-c.sol.1}
		\begin{split}		
W=\delta(\Delta\vp-\overline{\Delta\vp}),
\quad\sin(\overline{\Delta\vp}(A,\omega))=\frac{\omega-\Omega}{A}&, \ \
A\ge|\omega-\Omega|, \\
			W=\frac{C(A,\omega)}{|\omega-\Omega -
A\sin(\Delta\vp)|}&, \ \ A<|\omega-\Omega|.
		\end{split}
	\ee
	The first equation in~(\ref{s-c.sol.1}) has two solutions, we take the
microscopically stable one 
	\be
		\label{s-c.sol.2a}	
e^{\ii\overline{\Delta\vp}(A,\omega)}=\sqrt{1-\left(\frac{\omega-\Omega}{A}
\right)^2\,}+\ii\frac{\omega-\Omega}{A},
	\ee
	
	Also we need to calculate the following integrals, yielding the
contribution from the desynchronized oscillators
	\be
		\label{s-c.sol.3}
		\begin{split}
			C(A,\omega)=&\left(\int_0^{2\pi}
\frac{d\Delta\vp}{|\omega-\Omega -A
\sin(\Delta\vp)|}\right)^{-1}=\frac{\sqrt{(\Omega-\omega)^2-A^2}}{2\pi}\, , \\
			&\int_0^{2\pi}
\frac{e^{\ii\Delta\vp}\,d\Delta\vp}{|\omega-\Omega -A
\sin(\Delta\vp)|}=\frac{2\pi\ii}{A}\left(\frac{\Omega-\omega}{|\Omega-\omega|}
-\frac{\Omega-\omega}{\sqrt{(\Omega-\omega)^2-A^2}}\right) \, .
		\end{split}
	\ee
	Since in the integrals there is no dependence on $Q$, it is more
convenient to denote
	\be
		\label{s-c.Q.2}
		Q e^{\ii\Delta\Theta}=F(\W),
	\ee
	where
	\be
		\label{s-c.F.1}
		\begin{split}	
F(\W)&=\int_{|A|\ge|\Omega-\omega|}w(p)Be^{-\ii(\beta+\alpha)}\,
\sqrt{1-\frac{(\Omega-\omega)^2 }{A^2} \,}\, dp\, - \\
			&-\ii \int
w(p)Be^{-\ii(\beta+\alpha)}\,\frac{\Omega-\omega}{A}\, dp \, + \\
			&+\ii
\int_{|A|<|\Omega-\omega|}w(p)Be^{-\ii(\beta+\alpha)}\,\frac{\Omega-\omega}{
|\Omega-\omega|}\sqrt{\frac{(\Omega-\omega)^2 }{A^2}-1 \,}  \ dp\, .
		\end{split}
	\ee	
	Thus using relations~(\ref{s-c.Q.2})~and~(\ref{s-c.leader}) we obtain
the parametric solution of the problem:
	\be
		\label{s-c.Q.5}
		Q=|F(\W)|, \ \ \ \ \Delta\Theta=\text{arg}(F(\W)), \ \ \ \
{\omega_0}=\W-{\rm Im}(F(\W)).
	\ee

In the case of the Kuramoto-type model with generic
coupling described in~\cite{Vlasov-Macau-Pikovsky-14}, where a similar self-consistent
approach has been applied, the mean field has been characterized by two unknown variables,
the frequency $\Omega$ and the amplitude $Q$. The integral in
the function $F=F(\Omega,Q)$ was dependent on these variables, therefore 
two non-distributed parameters were needed in order to express them through the unknown variables.
In contradistinction, here for 
the leader-type coupling, only the frequency $\Omega$ enters the integral.
Thus, the solution here is parametrized
by the frequency of the leader $\Omega$ only, and thereby we have only one
non-distributed parameter of the original system that is expressed via the
complementary parameter $\W$, namely the natural frequency of the leader
$\omega_0$. So hereinafter we will represent the solutions in the form of the
dependence of $Q$ and $\Omega$ on the parameter $\omega_0$. Also the phase
$\Delta\Theta$ is not indicative, so we will not show it in the examples below.
		
In this model, the amplitude of the global field $Q$ that determines the forcing
acting on the oscillators is not normalized, and 
can be larger than unity. Moreover, it does not vanish for the asynchronous
regime. Thus it is not convenient to use it as an 
order parameter. As an order parameter it is more suitable  to use the relative
number of locked oscillators, or, in the 
thermodynamic limit, the parameter $R$ defined according to
Eq.~(\ref{s-c.orderparam}):
	\be
		\label{s-c.orderparam}
		R=\int_{|A|\ge|\Omega-\omega|}w(p)dp.
	\ee

%---------------------------------------------------------------

\subsection{Drums with a leader}

\begin{figure}[tbh]
		\includegraphics[width=0.5\columnwidth, clip]{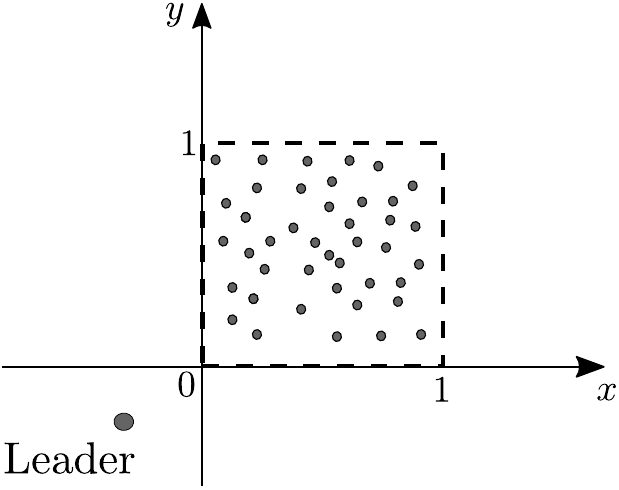}
		\caption{The scheme of the organization of the drum orchestra.}
		\label{fig.dr-sch}
	\end{figure}

Here, as an example of the application of this method, we will consider
system~(\ref{dr.gen.3}) as a model for the drum 
orchestra or any other  ensemble of oscillators with a spatial two-dimensional
organization. We assume that the
drum orchestra is a set 
of oscillators uniformly distributed on a unit square located at the
origin~(Fig.~\ref{fig.dr-sch}).
As in Ref.~\cite{Vlasov-Macau-Pikovsky-14}, where an example of  a geometric
organization of oscillators
has been treated, we assume that the phase shifts $\beta_j$ and $\alpha_i$ are
proportional to the distances between the oscillator and the leader, thus 	
	\be
		\label{ex.beta.1}
		\beta_i=\alpha_i={\omega_s \over c} \sqrt{(x_i-x_l)^2+(y_i-y_l)^2}\,,
	\ee
where $\omega_s$ is the central frequency of the original signal (around which
the phase approximation was made)
and $c$ is the speed of signal propagation.
Coupling strengths $B_j$ and $A_i$ are assumed to be inversely proportional to
the
square distances between each oscillator and the leader:
	\be
		\label{ex.B.1}
		B_j=\frac{W_B}{(x_j-x_l)^2+(y_j-y_l)^2},\quad
		A_i=\frac{W_A}{(x_i-x_l)^2+(y_i-y_l)^2}\;,
	\ee
	here additional initial intensities of the signals $W_A$ and $W_B$ were
added in order to have coupling coefficients of the order $1$ for any distant
position of the leader.
	
\begin{figure}[tbh]
		\includegraphics[width=0.45\columnwidth, clip]{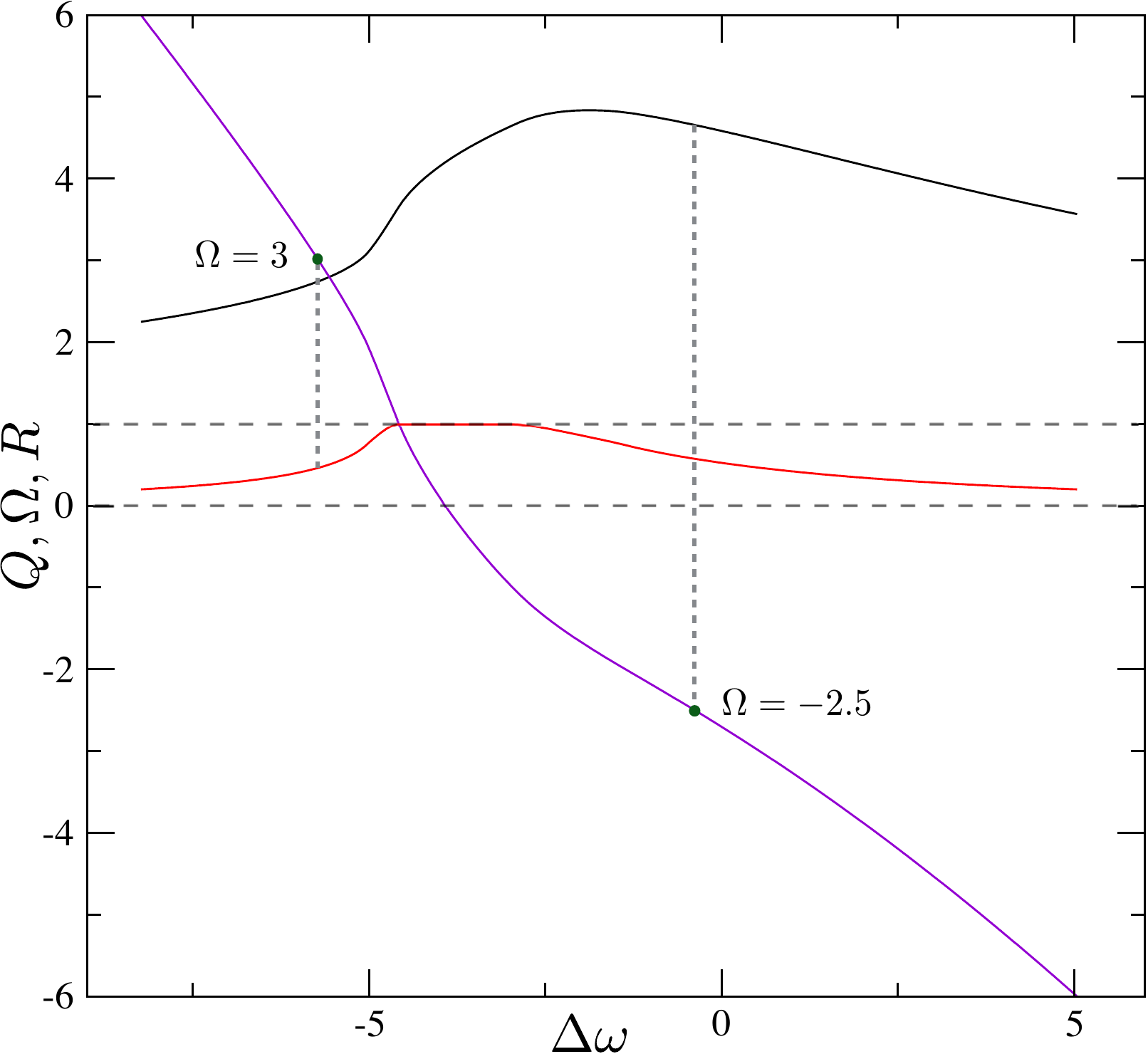}
		\caption{The dependences of the amplitude $Q$ (black curve) and
the frequency $\Omega$ (violet curve) of the global field together with the
order parameter $R$ (red curve) on the frequency mismatch
$\Delta\omega=\omega-\omega_0$, obtained self-consistently for the following
values of the parameters $x_l=-0.1$, $y_l=-0.1$ and $W_A=W_B$, $\omega_s/c=1$.
Horizontal dashed lines represent $1$ and $0$ on the y-axis.
Vertical lines represent the parameters for Fig.~\ref{fig.dr-3-phases}}
		\label{fig.dr-3}
	\end{figure}

Then in the thermodynamic limit the  distribution of the coupling parameters can be
written as $w(A, B, \alpha, \beta)=w(x,y)$, where 
all the parameters are the functions~(\ref{ex.beta.1},\ref{ex.B.1}) of the
coordinates $(x,y)$ of the 2D plane, except for, 
perhaps, natural frequencies $\omega$, that can be independently distributed. In
our numerical simulations we neglected 
this, assuming that all oscillators have identical frequencies. The
self-consistent approach gives solutions for any given 
position of the leader outside of the manifold of the oscillators, and for any
given value of its own natural frequency. 
As a measure of synchrony, we will use the order parameter $R$ introduced above
in Eq.~(\ref{s-c.orderparam}) (if $R$ 
is close to unity, the regime is synchronous, and if $R$ is small we call
this regime asynchronous). 
The terms ``synchronous'' and ``asynchronous'' are used here in order to show
the resemblance between the solutions of 
homogenous and non-homogenous systems. For the latter case, however, the usage of
these terms is not entirely correct as can 
be seen in Fig.~\ref{fig.dr-3}, where it is impossible to distinguish between
the partial synchrony and asynchrony because 
there is no abrupt transitions, and, except for a small region when all the
phases are locked ($R=1$), there is a fraction of 
locked phases and rotating phases with stationary distribution, that can be
named both as partial synchrony and as asynchrony 
in this case.

\begin{figure}[tbh]
		(a)\includegraphics[width=0.4\columnwidth, clip]{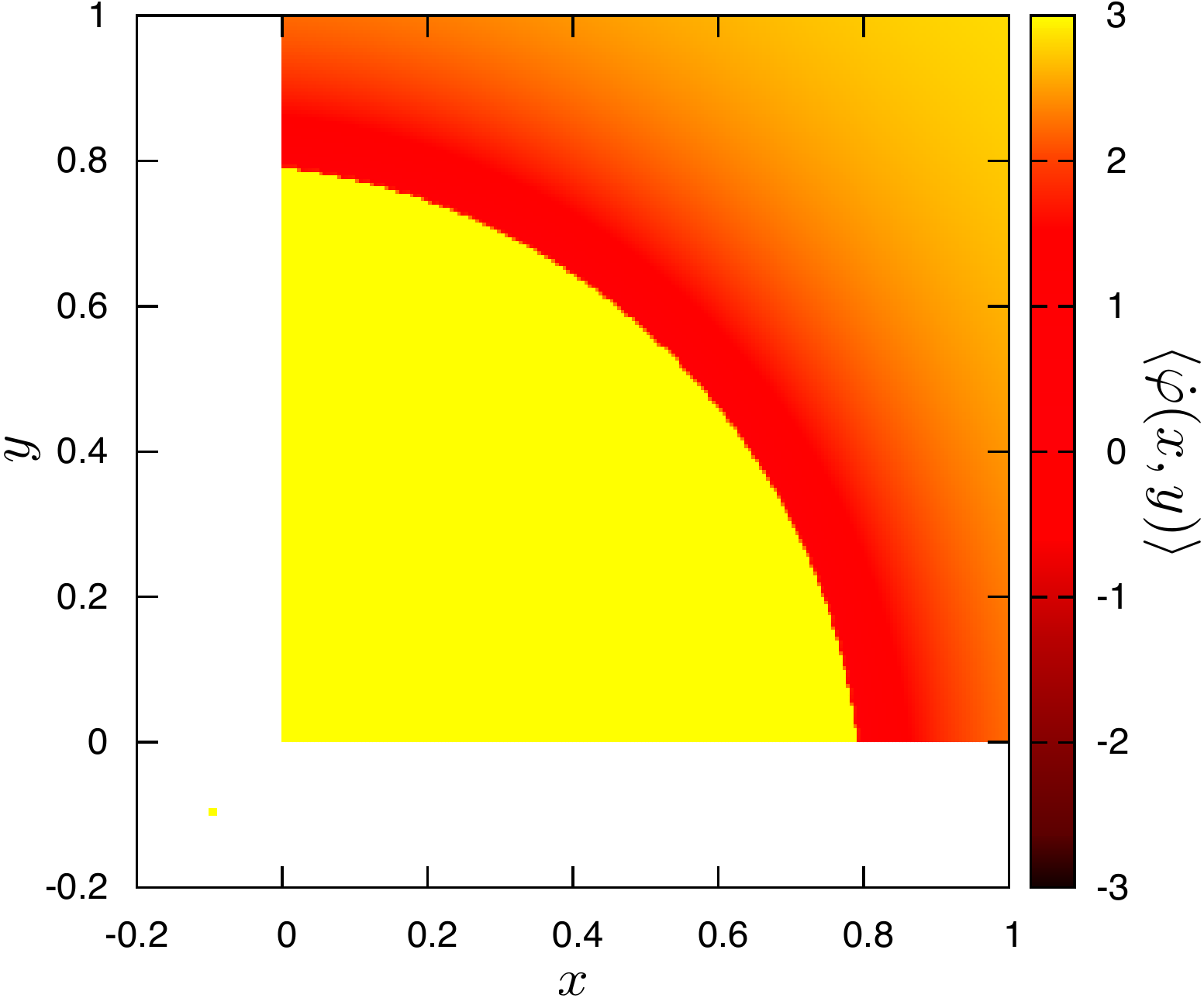}\hfill
		(b)\includegraphics[width=0.4\columnwidth, clip]{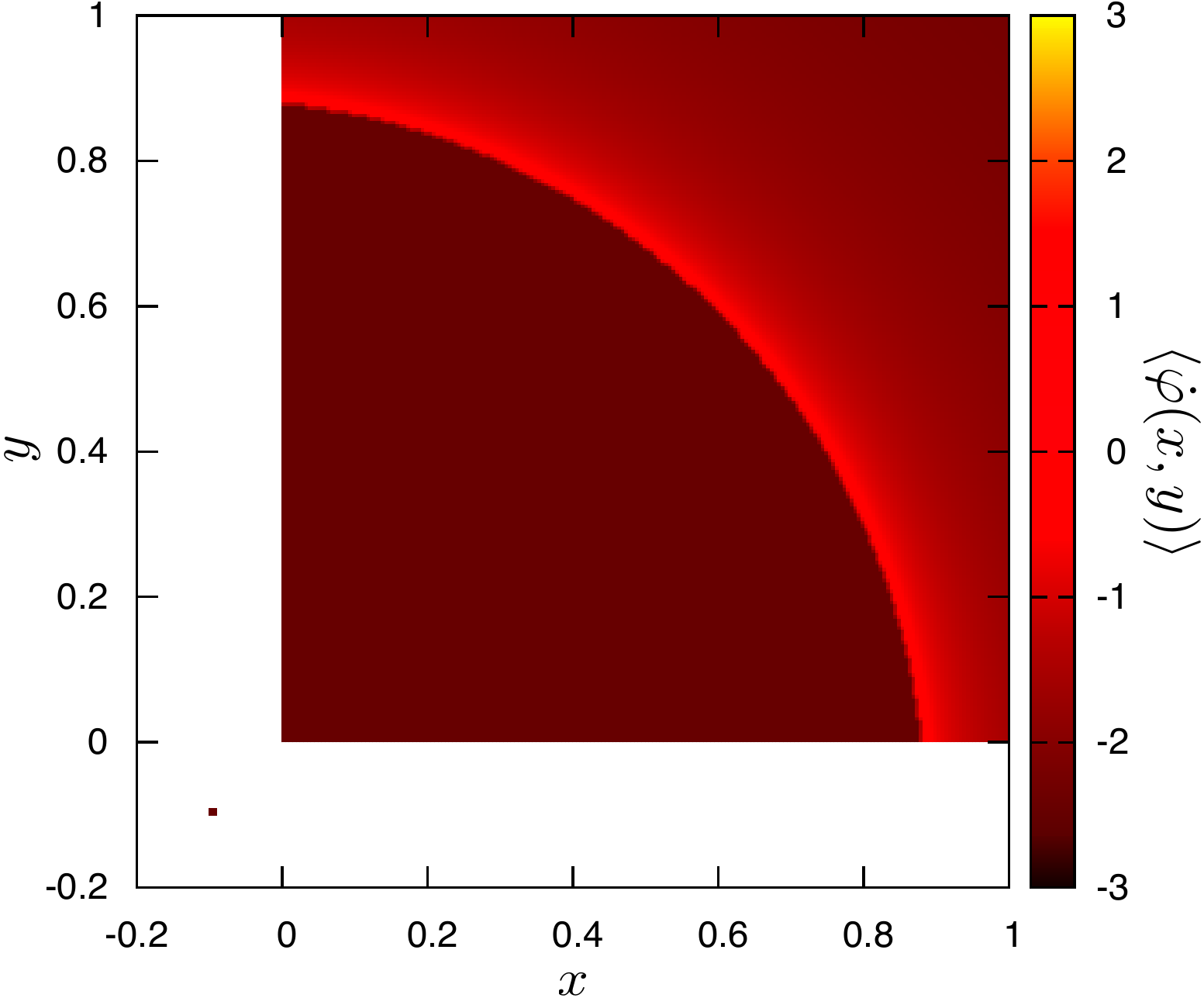}
		\caption{Distributions of the average frequencies of the oscillators for two cases
		(a) $\Omega=3$ and (b) $\Omega=-2.5$.
		For the locked phases $\dot\vp=\dot\phi=\Omega$ while
		for the rotating phases $\langle\dot\vp\rangle=\text{sign}(\Omega)\sqrt{(\Omega-\omega)^2-A^2}$.
		The parameters are the same as for Fig.~\ref{fig.dr-3}}
		\label{fig.dr-3-phases}
	\end{figure}

Spatial patterns of the average frequencies of the oscillators are shown on the Fig.~\ref{fig.dr-3-phases}
for positive and negative frequencies of the global field $\Omega$.
In both cases there is an area of locked phases, where $\dot\vp=\dot\phi=\Omega$, and an area of
rotating phases with $\langle\dot\vp\rangle=\text{sign}(\Omega)\sqrt{(\Omega-\omega)^2-A^2}$.

This model contains many parameters, varying each one can obtain various regimes. Therefore exhaustive description does not seem possible. While other complex states cannot be a priori excluded, next we will present the solutions that qualitatively coincide with the solutions of the homogenous system.

Fig.~\ref{fig.dr-3} shows the dependence of the amplitude $Q$, the frequency
$\Omega$ of the global field and the order parameter $R$ on the frequency
mismatch $\Delta\omega=\omega-\omega_0$ obtained self-consistently, for the case
when transitions between synchronous and asynchronous regimes are smooth. For
relatively large absolute values $|\Delta\omega|$, the order parameter $R$ is
small, but the value of the amplitude $Q$ of the global field is significantly
larger than zero. For  small negative $\Delta\omega$ the order parameter $R$
is equal to unity, meaning that all the phases are locked, what leads to even
larger values of $Q$. The interesting feature is that the maximum value of $Q$
is achieved when $R$ is smaller than unity. {The fully asynchronous
regime with $R=0$ is not shown on the Fig.~\ref{fig.dr-3}, however it exists for
sufficiently large absolute values of the frequency mismatch $|\Delta\omega|$.}
	
\begin{figure}[tbh]
		(a)\includegraphics[width=0.45\columnwidth, clip]{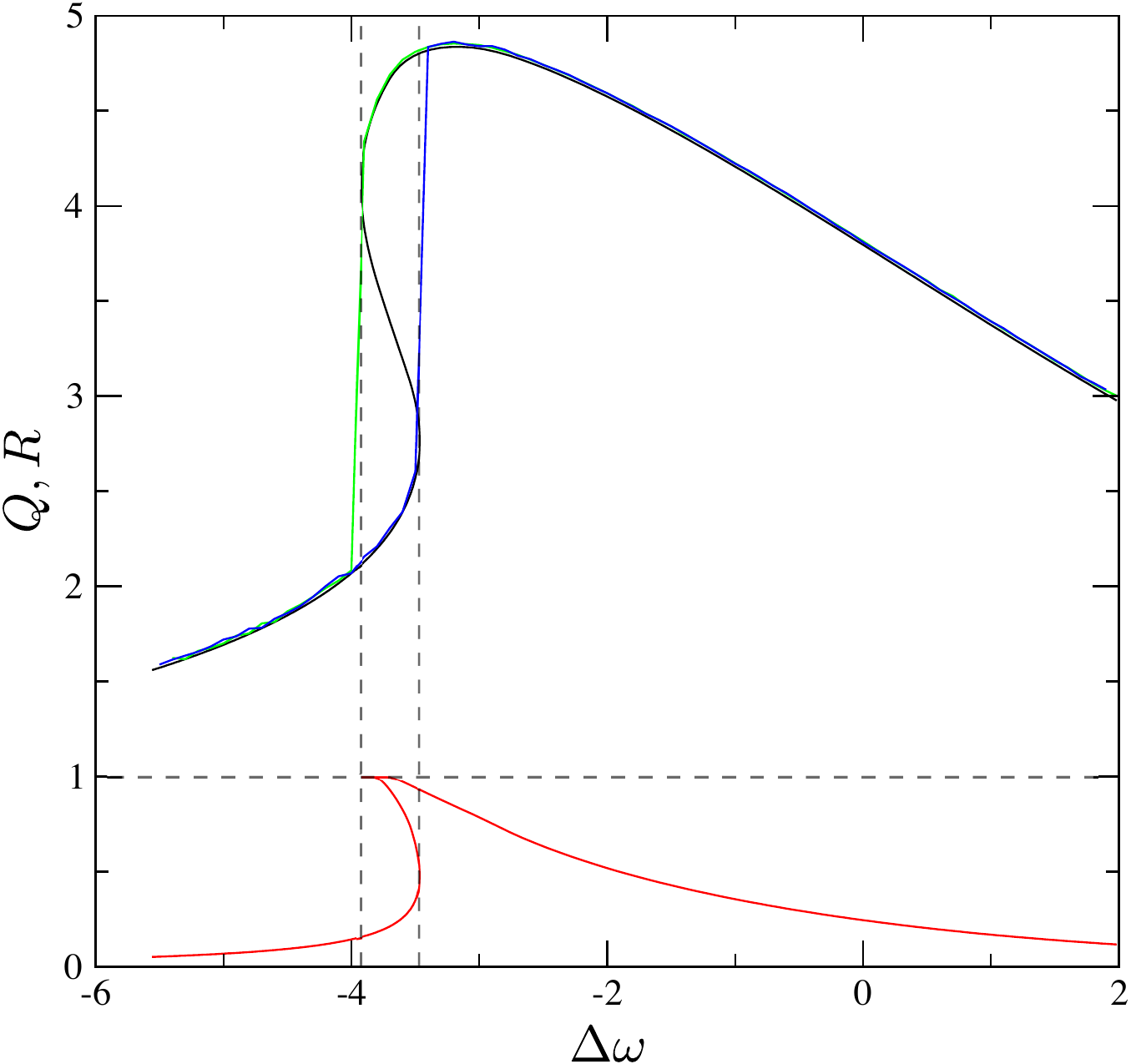} 
		(b)\includegraphics[width=0.45\columnwidth, clip]{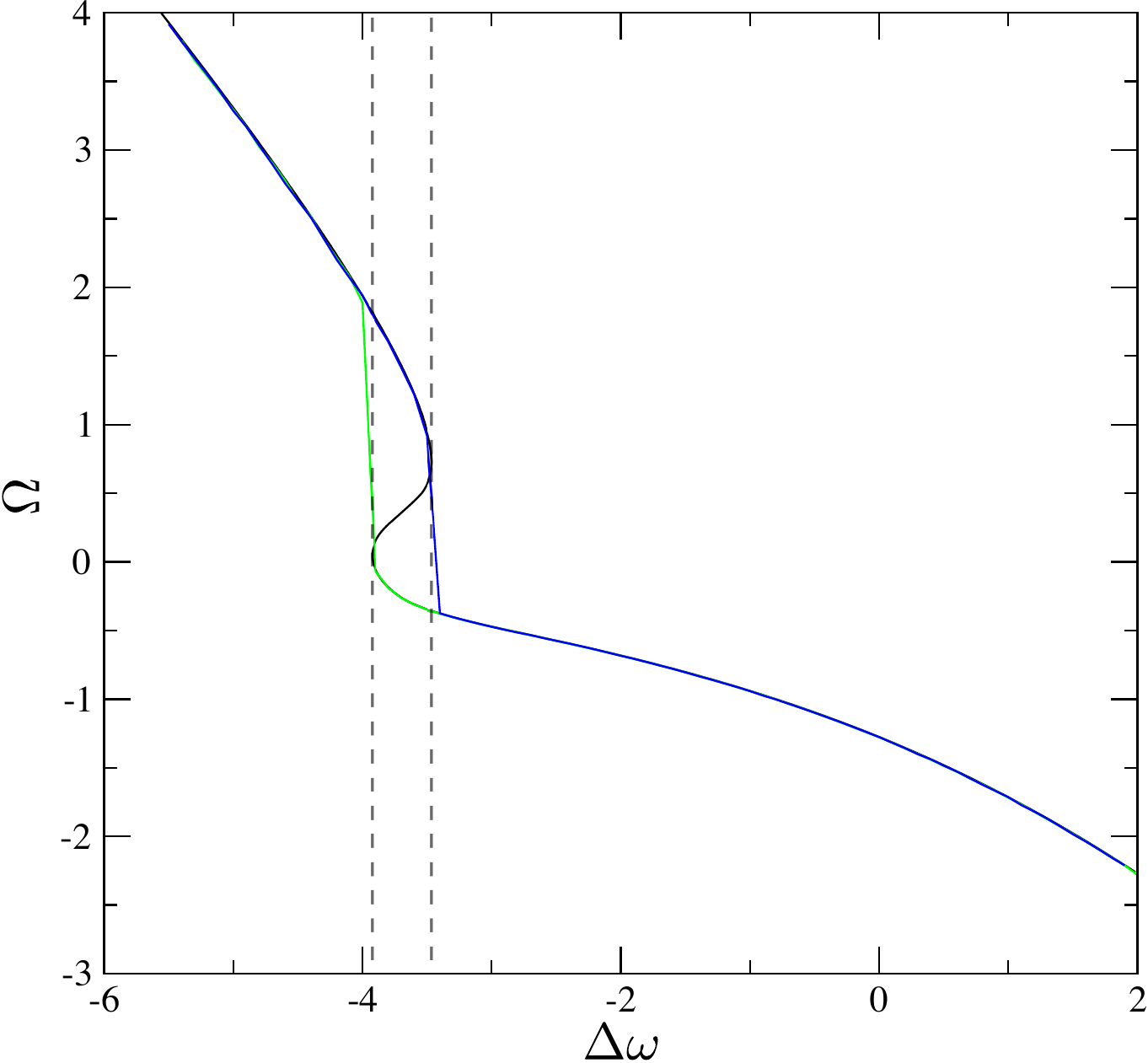}
		\caption{(a) The dependences of the amplitude $Q$ of
the global field (black curve represents self-consistent solution, blue and
green curves --- numeric)  and the order parameter $R$ (red curve obtained from
self-consistent method) as functions of the frequency mismatch
$\Delta\omega=\omega-\omega_0$, for the case $x_l=-0.1$, $y_l=-0.1$ and
$W_A=0.25 W_B$, $\omega_s/c=1$.
		(b) The dependences of the frequency $\Omega$ of the
global field (black curve from self-consistent approach is mostly hidden but can
be seen in the area of hysteresis, blue and green curves from numerics) as
functions of the frequency mismatch $\Delta\omega=\omega-\omega_0$, for the case
$x_l=-0.1$, $y_l=-0.1$ and $W_A=0.25 W_B$, $\omega_s/c=1$. {
Horizontal dashed line represents $1$ on the y-axis. Vertical dashed lines show
the area of hysteresis.}}
		\label{fig.dr-1-Q-Om}
	\end{figure}

In the second regime, we illustrate the situation when there are two stable
steady states (one asynchronous and one 
synchronous) with a hysteretic transition between them. The dependences of the
amplitude $Q$ and the 
frequency $\Omega$ of the global field on the frequency mismatch
$\Delta\omega=\omega-\omega_0$ for this case 
together with the order parameter $R$ are shown in  Fig.~\ref{fig.dr-1-Q-Om}. In
these figures we depict both the results 
obtained by the self-consistent method and by direct numerical simulations.
These two are very close (slight differences are due 
to the finite-size effects and the fact that we stop calculations at a finite
time) everywhere except for the area of the hysteresis 
that can be observed in the neighbourhood of the maximum amplitude of the global
field, and when the values of the order 
parameter $R$ are close to unity. For the large absolute values of the frequency
mismatch $|\Delta\omega|$, the order 
parameter $R$ is small {and goes to zero with further increase of
$|\Delta\omega|$}, what means that the number of locked phases is low and goes
to zero with further 
increasing of $|\Delta\omega|$. 

\begin{figure}[tbh]
		\includegraphics[width=0.45\columnwidth, clip]{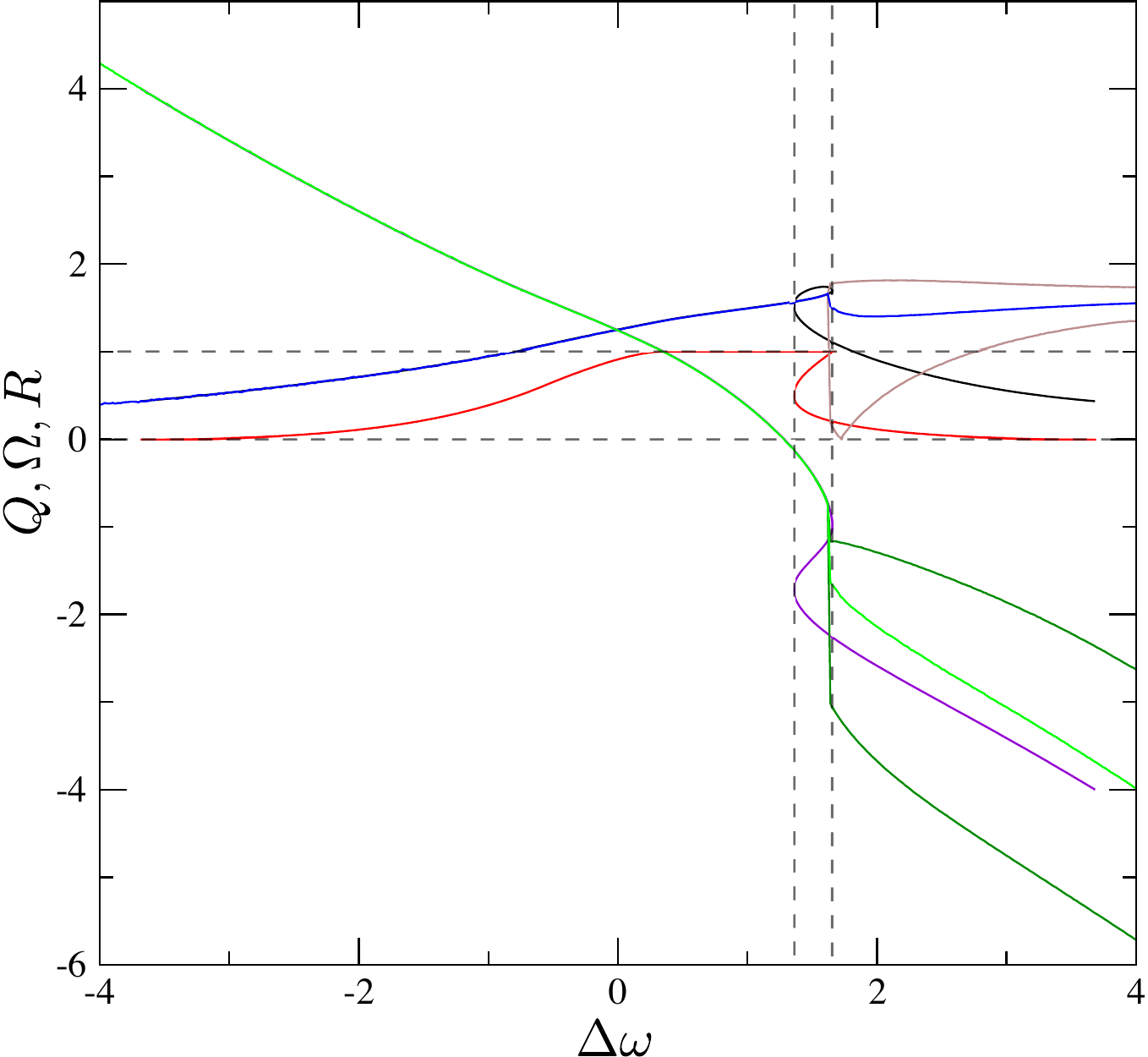}
		\caption{ The dependences of the amplitude $Q$ (black curve is
obtained self-consistently, blue curve is an average value on the limit cycle
obtained numerically and brown curves are minimum and maximum values on the limit
cycle) and the frequency $\Omega$ (violet curve --- self-consistent solution,
light green curve is numerical average over the limit cycle, dark green curves
are the minimum and the maximum on the limit cycle) of the global field on the
frequency mismatch $\Delta\omega=\omega-\omega_0$ together with the order
parameter $R$ (red curve obtained from self-consistent method). The following
values of the parameters were used $x_l=-1$, $y_l=-1$ and $W_A=W_B$,
$\omega_s/c=1$. {Horizontal dashed lines represent $1$ and $0$ on
the y-axis. Vertical dashed lines show the area of hysteresis.}}
		\label{fig.dr-2}
	\end{figure}

\begin{figure}[tbh]
		\includegraphics[width=0.45\columnwidth, clip]{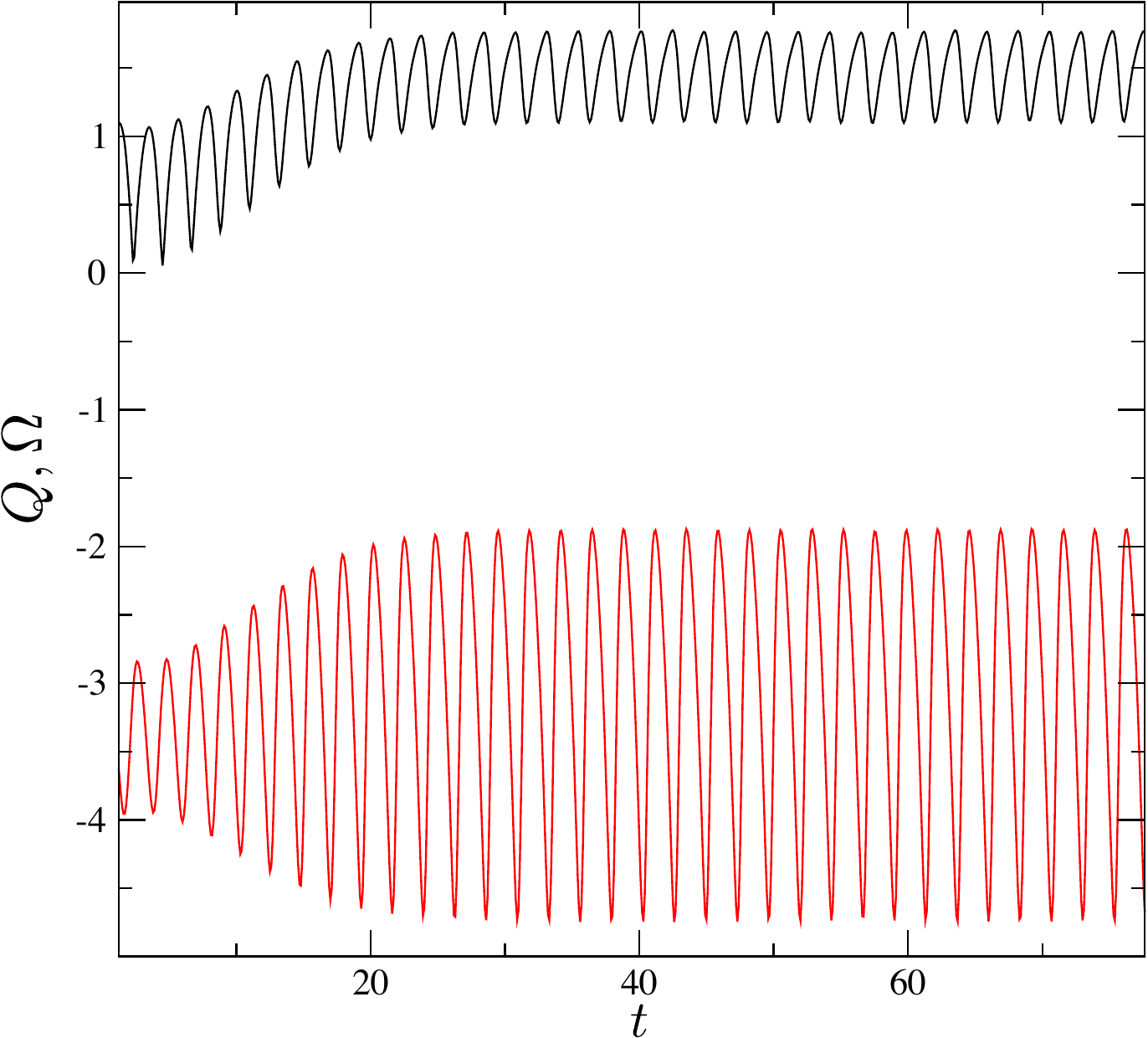}
		\caption{Numerical simulations for the amplitude $Q$ (black
curve) and the frequency $\Omega$ (red curve) of the global field as functions
of time for $\Delta\omega=\omega-\omega_0=3$, for the same values of the
parameters as in Fig.~(\ref{fig.dr-2}).}
		\label{fig.dr-2-t}
	\end{figure}

Another regime represents the case when there is one stable synchronous
fixed point and one unstable asynchronous fixed point. The results of numerical
simulations and the self-consistent method for this case are shown in
Fig.~\ref{fig.dr-2}. For negative and small positive values of
$\Delta\omega$, there is a steady solution. Being asynchronous for large negative
$\Delta\omega$, it gradually becomes partly synchronous for small negative
$\Delta\omega$, and transforms to a synchronous solution ($R=1$) for small positive
$\Delta\omega$. As can be seen with the help of numerical simulations, with
further increasing of $\Delta\omega$ the steady solution becomes unstable and we
observe the oscillating regime (these limit cycle oscillations are illustrated
in Fig.~\ref{fig.dr-2-t}).

In connection with the  interpretation of the model as a 
``drum orchestra'', the global field represents the signal
collected from all the drummers and filtered at the main frequency. This
frequency is equal to the frequency of the global field $\Omega$ and the
intensity of this filtered collected signal is equal to the amplitude of the
global field $Q$. The order parameter $R$ represents the relative amount of
drummers oscillating with the same main frequency of the collected signal
$\Omega$.
Our calculations show that perfect synchrony of the orchestra is not achieved
when the intensity of the filtered collected signal is maximum, therefore it
is not appropriate to use it as a measure for synchrony of the orchestra.

%-------------------------------------------------------------
\clearpage
\section{Conclusions}

In this paper we have performed the analysis of homogeneous and inhomogeneous
ensembles of oscillators coupled 
through a leader by virtue of two methods: the Watanabe-Strogatz approach for
identical oscillators, and the self-consistent theory 
for nonidentical oscillators. While the former method yields also stability of
the solutions, {in the second approach the stability 
properties are indicated with the help of direct numerical simulations}.
	
In the homogenous case, the WS approach gives a possibility for a full analytical
analysis of the system. The main result is the 
existence of a hysteretic transition from asynchrony to synchrony for some set of
the parameters, while for other sets the 
transition is not hysteretic. Another distinction from classical Kuramoto-type
systems is that the asynchronous solution 
has a non-zero order parameter.
	
In the inhomogeneous case, solutions rotating with a constant frequency have been
found self-consistently. As an example 
we consider the case when coupling parameters are determined by the spatial
distribution of the oscillators. In this situation the 
solutions similar to that of the homogeneous system can be found. In this case
the stability analysis cannot be performed, 
but with the help of direct numerical simulations we are able to show which
solutions are expected to be stable and which not.

Although our interpretation of the star-type network as a ``drum orchestra''
is sketchy and of course cannot be applied to real musical performances, the very field of 
acoustically coupled elements is, in our opinion, one where potential geometric
effects on synchronization due
to delay in signal propagation could be visible. Indeed, for a rhythm with a 1/2  second period,
propagation in air at distance of about 80 m would give a phase shift close to $\pi$, and
such distances are not unusual for large drumming sections.

The star-type network considered in this paper is just one example where the coupling
can be expressed through a mean field. More generally, there can be a set
of global variables (generalized mean fields) that act on oscillators. There are two main 
types of coupling: (i) the mean field is algebraically expressed through the states
of the oscillators, this is the case of the standard Kuramoto model and its generalization
to more complex coupling functions; (ii) the global variables are dynamical 
ones, i.e. there are dynamical equations for the mean fields. To this second
class belong situations where the global variables obey linear or nonlinear passive
equations (e.g., for Huygens clocks on a common support, the latter can be described
as a linear passive oscillator; similarly the common load for an array
of Josephson junctions is a passive linear or nonlinear LCR circuit). The star-coupled network
above is one where the global field is itself active, i.e. a self-sustained oscillator.
We studied only a situation where this central oscillator is of the same type as those in the battery, in particular
it is described by a similar phase equation. It would be interesting to study more general
setups, where the leader is, e.g., a weakly nonlinear self-sustained oscillator described
by the Stuart-Landau equations.
Another possible generalization is a system with several global fields. Partially we have touched it, when
both leader-mediated and direct Kuramoto-Sakaguchi-type coupling terms have been included.
Here of potential interest would be setups with two or more competing leaders, we expect that
the analytical approaches developed in this paper could be extended to studies of such problems.

\acknowledgments
We thank M. Komarov, C. Freitas, and M. Rosenblum for useful discussions.
V. V. thanks the IRTG 1740/TRP 2011/50151-0, funded by the DFG /FAPESP.  A.P. 
was supported by the grant according to the agreement of August 27,
2013 Nr 02.49.21.0003 between the Ministry of Education and Science of the Russian
Federation and Lobachevsky State University of Nizhni Novgorod. 
%------------------------------------------------------

\appendix
\section{Representing the order parameter in terms of Watanabe-Strogatz (WS)
global variables}
\label{App:id.WS.z}

	In order to represent the order parameter
	\be
		\label{WS.ordpar}
		Z={1\over N}\sum_{j=1}^N e^{\ii\vp_j}
	\ee	
	through the WS variables one should substitute the original phases
included as $e^{\ii\vp_k}$ with the WS transformation~(\ref{WS.1})
(see~\cite{Pikovsky-Rosenblum-11} for details), such that
	\be
		\label{WS.ordpar.1}
		Z={1\over N}\sum_{j=1}^N e^{\ii\vp_j}={1\over N}\sum_{j=1}^N
\frac{z+e^{\ii(\psi_j+\Psi)}}{1+z^*e^{\ii(\psi_j+\Psi)}}.
	\ee
	The expression~(\ref{WS.ordpar.1}) is rather complex and not applicable
for analytical analysis. But there is a special case when this expression
becomes extremely simple. First let us use an identity
	\be
		\label{WS.identity}
		\left(1+z^*e^{\ii(\psi_j+\Psi)}\right)^{-1}=\sum_{l=0}^\infty
(-z^*)^l e^{\ii l(\psi_j+\Psi)}.
	\ee
	Using the identity~(\ref{WS.identity}) we rewrite the
expression~(\ref{WS.ordpar.1}) for $Z$
	\be
		\label{WS.ordpar.2}
		Z={1\over N}\sum_{j=1}^N
\left(z+e^{\ii(\psi_j+\Psi)}\right)\sum_{l=0}^\infty (-z^*)^l e^{\ii
l(\psi_j+\Psi)},
	\ee
	or
	\be
		\label{WS.ordpar.3}
		Z=z\left[1+\left(1-|z|^{-2}\right)\sum_{l=1}^\infty (-z^*)^l
{1\over N}\sum_{j=1}^N e^{\ii l(\psi_j+\Psi)}\right],
	\ee
	
	Then, in the thermodynamic limit (the number of oscillators goes to
infinity), there is one special configuration of constants $\psi$ (the index has
been dropped because constants now have continuous distribution) when this
expression is simple. Such a configuration is a uniform distribution of
constants $\psi$. In this case the order parameter $Z$ is equal to the global
variable $z$ (due to the fact that the sums over $j$ in~(\ref{WS.ordpar.3})
become integrals over the distribution and in the case of the uniform
distribution these integrals vanish). Note that the requirement of the uniform
distribution of constants $\psi$ is a restriction on initial conditions, but it
does not mean that the initial conditions should be also uniformly distributed
(because $z(0)$ not necessary should be equal to zero).

%--------------------------------------------------------------

\section{Special case when $\cos\delta=-1$ and $B=A$}
\label{App:Special.case}

	If $\cos\delta=-1$ and $B=A$ then Eqs.~(\ref{id.WS.3}) transform to
	\be
		\label{App:Sc.id.WS.3}
		\begin{split}
			\dot{\rho}&=A\frac{1-\rho^2}{2}{\rm
Re}(e^{\ii\Delta\Phi}), \\	
\dot{\Delta\Phi}&=\Delta\omega-A\frac{1-\rho^2}{2\rho}\,{\rm
Im}(e^{\ii\Delta\Phi}).
		\end{split}
	\ee
	Thus in this special case, there are no synchronous steady states, only
limit cycle with $\rho=1$ and $\Delta\Phi(t)=\Delta\omega\,t$.
	The asynchronous steady states could be found from the equations
analogous to Eqs.~(\ref{l.id.WS.DeltaPhi=pi/2.1}), they read
	\be
		\label{App:Sc.l.id.WS.DeltaPhi=pi/2.1}
		\begin{split}
			\Delta\Phi&=\pm\,\pi/2, \\
			0&=\Delta \omega\mp A\frac{1-\rho^2}{2\rho}.
		\end{split}
	\ee
	Then the steady asynchronous solutions are
	\be
		\label{App:Sc.l.id.WS.DeltaPhi=pi/2.3.sol}	
{z_a}_{1,2}=\text{sign}(\Delta\omega)\,\ii\frac{\,-|\Delta\omega|\mp\sqrt{
\Delta\omega^2+A^2\,}\,}{A}.
	\ee
	From~(\ref{App:Sc.l.id.WS.DeltaPhi=pi/2.3.sol}) follows that
$|{z_a}_1|>1$ and $|{z_a}_2|<1$ if $\Delta\omega\neq 0$. And thus there is only
one asynchronous steady solution ${z_a}_2$. After linearization around ${z_a}_2$
the following linear system is obtained
	\be
		\label{App:Sc.l.id.WS.stab}
		\begin{split}		
\dot{a}&=-\,\text{sign}(\Delta\omega)\,\sqrt{\Delta\omega^2+A^2\,}\,b,\\
			\dot{b}&=\text{sign}(\Delta\omega)\,|\Delta\omega|a ,
		\end{split}
	\ee
	where $a={\rm Re}(z)$ and $b={\rm Im}(z)-{\rm Im}({z_a}_{2})$.
	Linear system~(\ref{App:Sc.l.id.WS.stab}) has two eigenvalues:
	\be
		\label{App:Sc.l.id.WS.stab.syn.eig}	
{\lambda_a}_2^{1,2}=\pm\sqrt{-|\Delta\omega|\sqrt{\Delta\omega^2+A^2\,}\,},
	\ee	
	what means that ${z_a}_2$ is neutrally stable as in the general case
when $\sin\delta=0$.	

%------------------------------------------------------
\section{The presentation of the solutions for all values of the parameters}
\label{App:id.sol}	

	Here we present a detailed description of the solution for all three
cases.
	Since we consider only the case when $\cos\xi>0$ (for $\cos\xi=0$ see
separate section) and thus $\sin\xi\neq1$, the solution with stability for
$\Delta x>0$ [$\Delta x<0$] is
	
	(i) (Fig.~\ref{fig.st-dr-123}(a)) $(2\sin\xi-g)>g$, note that $1>\sin\xi>
g\geq 0$ and
	\be
		\label{l.id.WS.stab.asyn.note.1}
		\left(|\Delta x|-(\sin\xi)\,\frac{|\Delta x|\pm\sqrt{\Delta
x^2-g(2\sin\xi-g)}\,}{2\sin\xi-g}\right)>0.
	\ee
	\be
		\label{l.id.WS.fp.st.1}
		\begin{split}
			&{z_s}_{1}-sink (stable) node \ [sink (stable) node] \ \
\ \text{and}\\
			&{z_s}_{2}-source (unstable) node \ [source (unstable)
node], \ \ \ \text{if}\ |\Delta x|<\sqrt{g(2\sin\xi-g)\,},\\
			\\
			&{z_a}_{1}-saddle \ [saddle], \ \ {z_a}_{2}-stable \
[unstable] \ \ \ \text{and} \\
			&{z_s}_{1}-sink (stable) node \ [sink (stable) node] \ \
\ \text{and} \\
			&{z_s}_{2}-source (unstable) node \ [source (unstable)
node], \ \ \ \text{if}\ \sqrt{g(2\sin\xi-g)\,}\leq|\Delta x|\leq\sin\xi, \\
			\\
			&{z_a}_{2}-stable \ [unstable] \ \ \ \text{and} \\
			&{z_s}_{1}-saddle \ [sink (stable) node]\ \ \ \text{and}
\\
			&{z_s}_{2}-source (unstable) node \ [saddle], \ \ \ 
\text{if}\ \sin\xi<|\Delta x|\leq 1, \\
			\\
			&{z_a}_{2}-stable \ [unstable], \ \ \ \text{and}\\
			&|z|=1, \ \text{arg}(z)=\Delta\Phi(t), \ unstable \
[stable] \ limit \ cycle \ \ \ \text{if}\ 1<|\Delta x|.
		\end{split}
	\ee

	(ii) (Fig.~\ref{fig.st-dr-123}(b)) $-g\leq(2\sin\xi-g)\leq g$, so
$g\geq\sin\xi\geq0$ and
	\be
		\label{l.id.WS.stab.asyn.note.2}
		\left(|\Delta x|-(\sin\xi)\,\frac{|\Delta x|\pm\sqrt{\Delta
x^2-g(2\sin\xi-g)}\,}{2\sin\xi-g}\right)>0
	\ee
	\be
		\label{l.id.WS.fp.st.2}
		\begin{split}
			&{z_s}_{1}-sink (stable) node \ [sink (stable) node] \ \
\ \text{and} \\
			&{z_s}_{2}-source (unstable) node \ [source (unstable)
node], \qquad \text{if}\ |\Delta x|<\sin\xi, \\
			\\
			&{z_a}_{2}-stable \ [unstable] \ \ \ \text{and} \\
			&{z_s}_{1}-saddle \ [sink (stable) node]\ \ \ \text{and}
\\
			&{z_s}_{2}-source (unstable) node \ [saddle], \qquad
\text{if}\ \sin\xi\leq|\Delta x|\leq 1, \\
			\\
			&{z_a}_{2}-stable \ [unstable], \ \ \ \text{and}\\
			&|z|=1, \ \text{arg}(z)=\Delta\Phi(t), \ unstable \
[stable] \ limit \ cycle \ \ \ \text{if}\ 1<|\Delta x|.		\end{split}
	\ee

	(iii) (Fig.~\ref{fig.st-dr-123}(c)) $(2\sin\xi-g)<-g$, thus $\sin\xi<0$ and
	\be
		\label{l.id.WS.stab.asyn.note.3}
		\begin{split}
			\left(|\Delta x|-(\sin\xi)\,\frac{|\Delta
x|+\sqrt{\Delta x^2-g(2\sin\xi-g)}\,}{2\sin\xi-g}\right)&<0 \\ 
			\left(|\Delta x|-(\sin\xi)\,\frac{|\Delta
x|-\sqrt{\Delta x^2-g(2\sin\xi-g)}\,}{2\sin\xi-g}\right)&>0	
\end{split}
	\ee
	\be
		\label{l.id.WS.fp.st.3}
		\begin{split}
			&{z_a}_{1}-unstable \ [stable], \ \ {z_a}_{2}-stable \
[unstable] \ \ \ \text{and} \\
			&{z_s}_{1}-saddle \ [saddle] \ \ \ \text{and} \\
			&{z_s}_{2}-saddle \ [saddle], \qquad \text{if}\ |\Delta
x|\leq|\sin\xi|, \\
			\\
			&{z_a}_{2}-stable \ [unstable] \ \ \ \text{and} \\
			&{z_s}_{1}-saddle \ [sink (stable) node]\ \ \ \text{and}
\\
			&{z_s}_{2}-source (unstable) node \ [saddle], \qquad
\text{if}\ |\sin\xi|<|\Delta x|\leq 1, \\
			\\
			&{z_a}_{2}-stable \ [unstable], \ \ \ \text{and}\\
			&|z|=1, \ \text{arg}(z)=\Delta\Phi(t), \ unstable \
[stable] \ limit \ cycle \ \ \ \text{if}\ 1<|\Delta x|.		
\end{split}
	\ee

	Where
	\be
		\label{l.id.WS.fp.note.s}
		\begin{split}
			&{z_s}_{1}=e^{\ii\left(\frac{\pi}{2}+\arcsin\Delta x
-\xi\right)},\\
			&{z_s}_{2}=e^{\ii\left(-\frac{\pi}{2}-\arcsin\Delta x
-\xi\right)},
		\end{split}
	\ee
	and if $2\sin\xi-g\neq 0$
	\be
		\label{l.id.WS.fp.note.as.1}
		\begin{split}
			&{z_a}_{1}=\text{sign}(\Delta x)\,\ii\frac{|\Delta
x|+\sqrt{\Delta x^2-g(2\sin\xi-g)}}{2\sin\xi-g}, \\
			&{z_a}_{2}=\text{sign}(\Delta x)\, \ii\frac{|\Delta
x|-\sqrt{\Delta x^2-g(2\sin\xi-g)}}{2\sin\xi-g},
		\end{split}
	\ee
	or if $2\sin\xi-g= 0$
	\be
		\label{l.id.WS.fp.note.as.2}
		\begin{split}
			&{z_a}_{1}=\text{sign}(\Delta x)\,\ii\frac{g}{2|\Delta
x|}, \\
			&{z_a}_{2}=-\,\text{sign}(\Delta
x)\,\ii\frac{g}{2|\Delta x|}.
		\end{split}
	\ee

%-------------------------------------------------------------

%

\end{document}